\documentclass[amsmath,amssymb,aps,prb,preprint,groupedaddress]{revtex4-2}

\usepackage{graphicx}
\usepackage{dcolumn}
\usepackage{bm}
\usepackage{ulem}
\usepackage{color,soul}

\usepackage{longtable}

\begin{document}


\title{Stochastic Fluid Dynamics Simulations of the Velocity Distribution in Protoplasmic Streaming }



\author{Vladislav Egorov \textit{$^{1}$}}
\author{Olga Maksimova \textit{$^{1}$}}
\author{Irina Andreeva \textit{$^{2}$}}
\author{Hiroshi Koibuchi \textit{$^{3}$}}\email[]{koi-hiro@sendai-nct.ac.jp; koibuchih@gmail.com}
\author{Satoshi Hongo \textit{$^{3}$}}
\author{Shinichiro Nagahiro \textit{$^{3}$}}
\author{Toshiyuki Ikai \textit{$^{3}$}}
\author{Madoka Nakayama \textit{$^{3}$}}
\author{Shuta Noro \textit{$^{3}$}}
\author{Tetsuya Uchimoto \textit{$^{4,5}$}}
\author{Jean-Paul Rieu \textit{$^{6}$}}
\affiliation{ $^{1}$\quad~Cherepovets State University (ChSU), 162600, Prospekt Lunacharskogo, 5, Cherepovets, Vologda Oblast, Russian Federation \\
$^{2}$\quad~Peter the Great St. Petersburg Polytechnic University (SPbPU), 195251, Polytechnicheskaya, 29, St.Petersburg, Russian Federation \\
$^{3}$\quad~National Institute of Technology (KOSEN), Sendai College, 48 Nodayama, Medeshima-Shiote, Natori-shi, Miyagi 981-1239, Japan \\
$^{4}$\quad~Institute of Fluid Science (IFS), Tohoku University, 2-1-1 Katahira, Aoba-ku Sendai 980-8577, Japan \\
$^{5}$\quad~ELyTMaX UMI 3757, CNRS - Universite de Lyon -Tohoku University, International Joint Unit, Tohoku University, 2-1-1 Katahira, Aoba-ku Sendai 980-8577, Japan \\
$^{6}$  Univ Lyon, Universit$\acute{e}$ Claude Bernard Lyon 1, CNRS UMR-5306, Institut Lumi$\grave{e}$re Mati$\grave{e}$re, F-69622, Villeurbanne, France
}



\begin{abstract}
Protoplasmic streaming in plant cells is directly visible in the cases of \textit{Chara corallina} and \textit{Nitella flexilis}, and this streaming is understood to play a role in the transport of biological materials. 
For this reason, related studies have focused on molecular transportation from a fluid mechanics viewpoint. However, the experimentally observed distribution of the velocity along the flow direction $x$, which exhibits two peaks at $V_x\!=\!0$ and at a finite $V_x(\not=\!0)$, remains to be studied. In this paper, we numerically study whether this behavior of the flow field can be simulated by a 2D stochastic Navier-Stokes (NS) equation for Couette flow, in which random Brownian force is assumed. We present the first numerical evidence that these peaks are reproduced by the stochastic NS equation, which implies that the Brownian motion of the fluid particles plays an essential role in the emergence of these peaks in the velocity distribution. We also find that the position of the peak at $V_x(\not=\!0)$ moves with the variation in the strength $D$ of the random Brownian force, which also changes depending on physical parameters such as the kinematic viscosity, boundary velocity and diameter of the plant cells.
\end{abstract}
%

\maketitle


\section{Introduction\label{intro}}
A circular flow called protoplasmic streaming is directly observed in the cells of specific plants, such as \textit{Chara corallina} and \textit{Nitella flexilis}, in which the cell size is relatively large, ranging from a few hundred micrometers to one millimeter \cite{Kamiya-1986,Kamiya-Kuroda-1973,Tazawa-pp1968,Kamiya-Kuroda-1958,Kamiya-Kuroda-1956}. Such streaming inside cells is naturally considered to play a role in the transportation of biological materials.

The driving force of the flow is known to be molecules moving along actin filaments; hence, these molecules are called molecular motors \cite{Houtman-etal-EPL2007,Klumpp-etal-PhysE2005,Lypowsky-etal-PhysA2006,Kawakubo-etal-PhysA1998,Julicher-etal-RMP1997,Astumianl-Science1997}. These molecular motors transport chlorophyll, which is very large, and drive the flow in plant cells. Interestingly, the speed of the flow in cells is closely related to the size of the plant \cite{Tominaga-etal-DevCell2013}. Moreover, the mechanism for the transportation of biological materials is understood to be the same as that in animal cells \cite{Tominaga-Ito-CuOpPB2015,McIntosh-Ostap-JCS2016}.

\begin{figure}[t]
\begin{center}
\includegraphics[width=9.5cm]{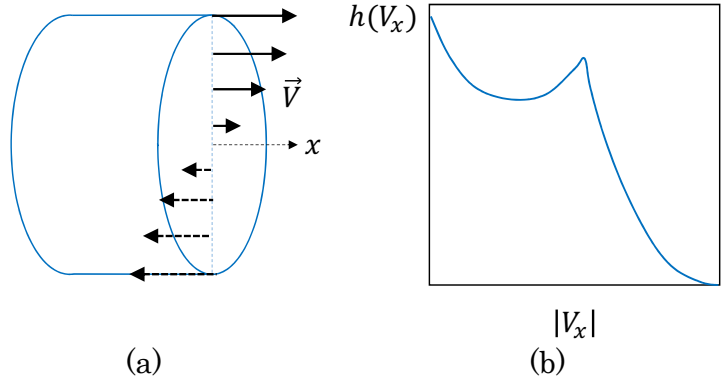}
\caption{ (a) The flow velocity ${\vec V}$ inside a cell. (b) The normalized velocity distribution $h(V_x)$ along the $x$-direction. }
 \label{fig-1}
\end{center}
\end{figure}

Therefore, protoplasmic streaming has attracted substantial attention both in scientific fields and in the context of agricultural technology \cite{Tominaga-etal-DevCell2013}. Kamiya and Kuroda observed the position dependence of the flow speed in a section vertical to the longitudinal direction of \textit{Nitella} cells via optical microscopy in 1956 \cite{Kamiya-Kuroda-1956} (Fig. \ref{fig-1}(a)).
This position dependence of the flow speed was later precisely measured via particle tracking velocimetry by Kikuchi-Mochizuki \cite{Kikuchi-Mochizuki-PlosOne2015}, who reported results compatible with the simulation data obtained by Goldstein et al. using coupled Navier-Stokes (NS) and advection-diffusion equations \cite{Goldstein-etal-PRL2008,Goldstaein-etal-PNAS2008}. Goldstein et al. assumed a spiral flow as a boundary condition on the wall in their simulations and provided insight into the role of this spiral flow in molecular transportation. This velocity field was later shown to be compatible with experimental data obtained through magnetic resonance velocimetry \cite{Goldstaein-etal-JFM2010,Raymond-Goldstein-IF2015}. Niwayama et al. also simulated streaming in the case of \textit{Chara corallina} using the 3D NS equation in a method called the moving particle semi-implicit method, in which the spiral flow is neglected, and reported results almost identical to those of Goldstein et al. \cite{Niwayama-etal-PNAS2010}. Their original motivation was to use particle image velocimetry to measure the streaming velocity in the case of \textit{Caenorhabditis elegans} embryos, in which the mechanism of transportation is slightly different from that in the cases of \textit{Chara corallina} and \textit{Nitella flexilis} \cite{Niwayama-etal-PNAS2010}.

\begin{figure}[ht]
\begin{center}
\includegraphics[width=9.5cm]{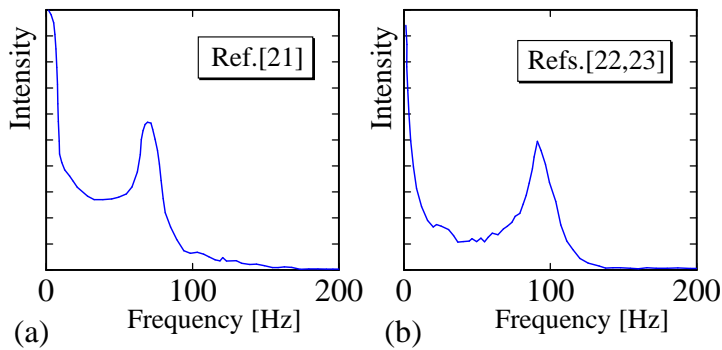}
\caption{   Plots of experimental velocity distributions obtained via laser Doppler velocimetry in (a) Ref. \cite{Mustacich-Ware-PRL1974} and (b) Refs. \cite{ Mustacich-Ware-BJ1976,Mustacich-Ware-BJ1977}. The horizontal axis represents the frequency of the laser light, which is proportional to the fluid velocity. 
 \label{fig-2} }
\end{center}
\end{figure}

In 1974, Mustacich and Ware observed the distribution of the velocity $V_x$ along the flow direction $x$ by means of a laser light scattering technique called laser Doppler velocimetry and found two different peaks in the velocity distribution: one at $V_x\!=\!0$ and one at a finite $V_x(\not=\!0)$ \cite{Mustacich-Ware-PRL1974,Mustacich-Ware-BJ1976,Mustacich-Ware-BJ1977} (Fig. \ref{fig-1}(b)). Shortly thereafter, the velocity distribution was again measured using the same technique by Sattelle and Buchan \cite{Sattelle-Buchan-JCS1976}, who similarly detected two different peaks.
Figures \ref{fig-2}(a) and \ref{fig-2}(b) show experimental data extracted from Refs. \cite{Mustacich-Ware-PRL1974,Mustacich-Ware-BJ1976,Mustacich-Ware-BJ1977}, where the data are represented by solid lines approximating the data points. The horizontal axis represents the frequency of the laser light, which is proportional to the fluid velocity, and the shape and position of the second peak depend on the scattering angle; note that the corresponding velocities in the figures are identical to each other for data obtained at the same point inside the cell. The velocity of the second peak was reported to be $60~{\rm \mu m/s}$ \cite{Mustacich-Ware-PRL1974} and $72~{\rm \mu m/s}$ \cite{Mustacich-Ware-BJ1976,Mustacich-Ware-BJ1977} for the data shown in Figs. \ref{fig-2}(a) and \ref{fig-2}(b), respectively. The biological implications of the existence of the second peak are currently unclear; however, it is possible that the peak in a relatively high-velocity region may be closely related to an enhancement of some biological function, such as transportation or mixing.

However, the experimentally obtained velocity distribution has not yet been numerically verified. Although the peak at zero velocity is expected to be caused by the Brownian motion of the fluid molecules, as noted in Ref. \cite{Sattelle-Buchan-JCS1976}, the peak at a finite velocity has yet to be explained. Clearly, a microscopic perspective is effective for studying this problem; therefore, to this end, we adopt Langevin simulation, which is a technique for simulating the Brownian motion of small particles \cite{Lemos-AmJP1997,Wilhelm-Springer1989,Metzler-PhysRep2000,KGWilson-PRD1985,Ukawa-Fukugita-PRL1985,Hofler-Schwarzer-PRE2000,Koibuchi-JPG1987,Nagahiro-PRE2007}.

The peak in the velocity distribution at zero velocity can be naturally understood from the fact that the fluid at the central part of a cell is expected to 
have a slow speed compared to the fluid at the wall \cite{Goldstein-etal-PRL2008,Goldstaein-etal-PNAS2008}; thus, the fluid in the central region is expected to be influenced by random Brownian forces.
In contrast, the fluid close to the cell wall is strongly influenced by the activation forces of molecular motors; in other words, thermal fluctuations are suppressed by contact with the motors and the cell wall. On the other hand, fluid that is separated from the cell wall is not influenced by such boundary conditions, and the speed of the fluid is expected to continuously decrease toward the central region of the cell. Therefore, no intuitive explanation is available for the existence of the second peak at a relatively high velocity.

In this paper, we numerically solve the NS equation with random Brownian force for flow fields in a square region by regarding twisting flows as straight flows along the longitudinal direction. This 2D NS equation is considered a Langevin equation or a stochastic differential equation because it includes a random force. In this paper, we combine two different techniques \cite{Uma-etal-PhysFluids2011}: NS simulation for continuum fluids \cite{Taylor-NSeq-2010,Lukaszewicz-2015} and Langevin simulation for particles \cite{Lemos-AmJP1997,Wilhelm-Springer1989,Metzler-PhysRep2000,KGWilson-PRD1985,Ukawa-Fukugita-PRL1985,Hofler-Schwarzer-PRE2000,Koibuchi-JPG1987,Nagahiro-PRE2007}.  This simulation approach in combination with the NS equation is new and, thus, is not comparable to standard techniques for the NS equation without random forces; therefore, we carefully check the dependence of the results on parameters including spatial and temporal discretizations. 

 It will be shown that all qualitatively different simulation results can be obtained by merely varying the strength $D$ of the random Brownian force and that two different peaks appear in the velocity distribution at intermediate values of $D$.

\section{Stochastic Navier-Stokes equation\label{SNS-equation}}
The symbols for the variables and constants used in this paper are listed along with their units and descriptions in Table \ref{table-1}.
\noindent
\begin{longtable}{ccp{100mm}}
\caption{List of symbols with units and descriptions. The numbers inside the parentheses are the assumed typical values used as inputs to the simulations. }
  \label{table-1} \\
  \hline
Symbol  & Unit & Description (assumed typical value)  \\ \hline
\endfirsthead
\multicolumn{3}{r}{continue} \\ \hline
Symbol  & Unit & Description (assumed typical value)  \\ \hline
\endhead
  \hline
\multicolumn{3}{r}{continue} \\
\endfoot
  \hline
\multicolumn{3}{r}{end} \\
\endlastfoot
$\psi$ & ${\rm m^2/s}$       &  Stream function   \\
$\omega$ & ${\rm 1/s}$      &  Vorticity    \\
$\vec V\!=\!(V_x,V_y)$ & ${\rm m/s}$      &  Velocity vector  \\
$V_B$ & ${\rm m/s}$      &  Velocity at the boundary  ($50\!\times\!10^{-6}$) \\
${\vec \eta}\!=\!(\eta_x,\eta_y)$ & ${\rm m/s^2}$      &  Gaussian random force   \\
${\vec g}\!=\!(g^x,g^y)$ & $1$      &  Gaussian random number of mean 0 and deviation 1  \\
$D$ & ${\rm m^2/s^3}$      &  Strength of random force  ($2\!\times\!10^{-7}\leq D\leq 4\!\times\!10^{-6}$) \\
$D_{\rm dif}$ & ${\rm m^2/s}$      &  Diffusion constant   \\
$\tau_e$ & ${\rm s}$      &  Macroscopic relaxation time   \\
$\mu$ & ${\rm Ns/m^2}$      &  Viscosity    \\
$\nu$ & ${\rm m^2/s}$      &  Kinematic viscosity ($1\!\times\!10^{-4}$)  \\
$n_X$ & $1$      &  Total number of lattice points on one edge ($100\leq n_X\leq 300$)  \\
$n_T$ & $1$      &  Total number of iterations per time step  \\
$\alpha$ & $1$      &  Positive number for a unit change in length \\
$\beta$ & $1$      &  Positive number for a unit change in time \\
$\gamma$ & $1$      &  Positive number for a change in $n_X$ \\
$\delta$ & $1$      &  Positive number for a change in $n_T$ \\
$d$ & ${\rm m}$      &  Cell diameter ($500\!\times\!10^{-6}$)\\
${\it \Delta}x$ & ${\rm m}$      &  Lattice spacing \\
${\it \Delta}t$ & ${\rm s}$      &  Discrete time step ($8\!\times\!10^{-9}$) \\
$E$ &       &  Parameter set $(\nu,  V,   D)$  \\
$S$ &       &  Parameter set $(\nu,  V,   D, {\it \Delta} x, {\it \Delta} t)$  \\
$a$ &  ${\rm m}$  &  Diameter of a lump of fluid particles  \\
${\rm Exp}(E)$ &       &  Experimental data corresponding to $E\!=\!(\nu,  V,   D)$  \\
${\rm R_e}$ &  $1$     &  Reynolds number  \\
${\rm P_e}$ &  $1$     &  P${\rm \acute e}$clet number  \\
${\rm S_c}$ &  $1$     &  Schmidt number  \\
\end{longtable}

\subsection{Discretization of the stochastic Navier-Stokes equation for the stream function \label{discretization}}%

\begin{figure}[t]
\begin{center}
\includegraphics[width=11.5cm]{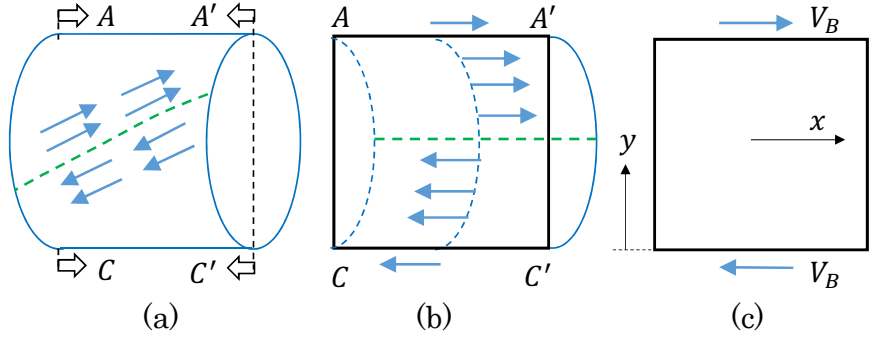}
\caption{
(a) Flow directions in the so-called indifferent zone on the surface and section $AA^\prime\!-\! CC^\prime$ of the cylinder at the center. (b) Section $AA^\prime\!-\! CC^\prime$ and the flow directions on the surface of the cylinder. (c) 2D simulation domain corresponding to section $AA^\prime\!-\! CC^\prime$, where $V_B$ denotes the fixed velocity considered as a boundary condition. }
 \label{fig-3}
\end{center}
\end{figure}

First, we present the computational domain extracted from the cylindrical body shown in Fig. \ref{fig-3}(a). The arrows in Fig. \ref{fig-3}(a) illustrate the flow directions in the indifferent zone indicated by the dashed line on the surface. We consider the section $AA^\prime-CC^\prime$, which is also visualized in Fig. \ref{fig-3}(b), with arrows on the boundaries $AA^\prime$ and $CC^\prime$, where the flow direction is modified to be parallel/antiparallel along the longitudinal direction. The arrows inside the square represent the velocity on the surface of the opposite side, which is not included in or differs from the square domain. The computational domain is the flat square region with boundaries $AA^\prime$ and $CC^\prime$, where the other two boundaries, $AC$ and $A^\prime C^\prime$, are assumed to be periodic. This square region is extracted and shown in Fig. \ref{fig-3}(c) to clarify the 2D nature of the domain. 

The boundary condition given by the velocity $V_B$ is simply the same as that for Couette flow. The real 3D flow is modified to this 2D flow for simplicity, and the flow direction in the 2D domain is obtained by modifying the flow direction on the surface of the cylinder, as stated above. Because of this modification of the velocity direction, we can determine whether the origin of the peaks in the velocity distribution lies in the spiral flow. It is also possible to investigate whether the peaks are related to the 3D nature of the flow.

Here, we should comment on the reason why the shape of the boundary $AA^\prime\!-\! CC^\prime$ is assumed to remain unchanged. Cell surfaces composed of soft biological materials may exhibit shape deformations that can be directly measured, for example, in the case of Caenorhabditis elegans embryos, in which cytoplasmic streaming is also expected \cite{Niwayama-etal-PNAS2010}. However, in the case of plants such as \textit{Chara corallina} and \textit{Nitella flexilis}, the situation is different; the cell surface is relatively hard, and fluctuations can be neglected.

The continuous form of the NS equation \cite{Taylor-NSeq-2010,Lukaszewicz-2015} with random Brownian force is given by
\begin{eqnarray}
\label{NS-eq}
\begin{split}
&\frac {\partial \omega}{\partial t}=-\left ({\vec V}\cdot \nabla\right)\omega+\nu {\it \Delta} \omega + \left(\nabla \times {\vec \eta}(t)\right)_z, \\
&\omega=-{\it \Delta} \psi,
\end{split}
\end{eqnarray}
where ${\vec V}\!=\!(V_x,V_y,0)$ is the fluid velocity obtained from the stream function $\psi$ and $\omega$ is the third component of the vorticity $\vec \omega\!=\!\nabla \times {\vec V}$, such that
\begin{eqnarray}
\label{stream-function}
\begin{split}
&V_x=-\frac {\partial \psi}{\partial y}, \quad V_y=\frac {\partial \psi}{\partial x}, \quad V_z=0\\
&\vec \omega=\nabla \times {\vec V}=(0,0,\omega).
\end{split}
\end{eqnarray}
The physical meaning of each term in Eq. (\ref{NS-eq}) will be given below in reference to the equation for the velocity $\vec V$.
The parameter $\nu$ in Eq. (\ref{NS-eq}) is the kinematic viscosity coefficient, where $\nu\!\simeq\!1\!\times\!10^{-6}~{\rm m^2/s}$ in the case of water at room temperature. The symbol ${\vec \eta}(t)\!=\!(\eta^x,\eta^y,0)$ represents Gaussian white noise or a Gaussian random force corresponding to the Brownian motion of the fluid particles or a lump of fluid particles. The components of ${\vec \eta}(t)$ are assumed to satisfy
\begin{eqnarray}
\label{Gaussian-random-force}
\langle \eta_i^\mu(t)\eta_j^\nu(t^\prime)\rangle = 2D\delta_{ij}\delta^{\mu \nu}\delta(t-t^\prime),
\end{eqnarray}
where $\langle \cdots\rangle$ denotes the expectation value, $D$ is called the strength of the random force and the subscript $i$ denotes the fluid position. In Eq. (\ref{Gaussian-random-force}), we introduce ${\vec \eta}(t)$ in a discrete form because the NS equation is discretized on a square lattice in this numerical study. No confusion is expected between the symbol for the kinematic viscosity coefficient $\nu$ and the superscript of the Gaussian random force $\eta^\nu(t)$.

Here, we comment on the reason why the NS equation for the stream function is used instead of the NS equation for the velocity field. Indeed, it is easy to check that Eq. (\ref {NS-eq}) is obtained from the following NS equation:
\begin{eqnarray}
\label{NS-eq-org-0}
\rho \left[\frac{\partial {\vec V}}{\partial t}+\left ({\vec V}\cdot \nabla\right){\vec V}\right]=- \nabla p +\mu {\it \Delta} {\vec V} +{\rho} {\vec \eta}(t),
\end{eqnarray}
where $\rho$ and $p$ are the density and pressure of the fluid, respectively, and $\mu\!=\!\rho\nu$ is the viscosity.
The fluid is assumed to be Newtonian. The NS equation in Eq. (\ref{NS-eq-org-0}) has the form of the standard equation of motion per unit volume for a fluid of density $\rho$. The second term on the left-hand side (LHS), which is called the advection term, arises from the fact that the fluid particles are moving with velocity $\vec V$. This term is very small compared with the other terms in the case of protoplasmic streaming; however, we include it for completeness. The first term on the right-hand side (RHS) represents the force from the pressure $p$; the negative sign appears by definition. The second term on the RHS, defined by the Laplace operator, represents the force from the viscosity of the fluid. The final term on the RHS is the random Brownian force $\eta(t)$, determined by Gaussian random numbers, on which detailed information will be given below.

Equation (\ref{NS-eq-org-0}) can be conveniently modified by multiplying both sides of the equation by $\rho^{-1}$, and by additionally incorporating the condition $\nabla\cdot {\vec V}\!=\!0,$ we obtain
\begin{eqnarray}
\label{NS-eq-org}
\begin{split}
&\frac{\partial {\vec V}}{\partial t}=-\left ({\vec V}\cdot \nabla\right){\vec V}-{\rho}^{-1}  \nabla p +\nu {\it \Delta} {\vec V} + {\vec \eta}(t),\\
&\nabla\cdot {\vec V}=0.
\end{split}
\end{eqnarray}
By multiplying this standard NS equation from the left by the rotation $\nabla\times$, we obtain the NS equation in Eq. (\ref{NS-eq}).
The NS equation in Eq. (\ref {NS-eq}) is used instead of Eq. (\ref{NS-eq-org}) because the condition $\nabla\cdot {\vec V}\!=\!0$ is exactly satisfied in Eq. (\ref {NS-eq}); therefore, Eq. (\ref {NS-eq}) is easier to solve numerically than the original NS equation for ${\vec V}$ given in Eq. (\ref{NS-eq-org}) for the case of protoplasmic streaming \cite{Hofler-Schwarzer-PRE2000}. This is the reason why the NS equation for the stream function, i.e., Eq. (\ref{NS-eq}), is used instead of the NS equation for the velocity field, i.e., Eq. (\ref{NS-eq-org}).

\begin{figure}[t]
\begin{center}
\includegraphics[width=9.5cm]{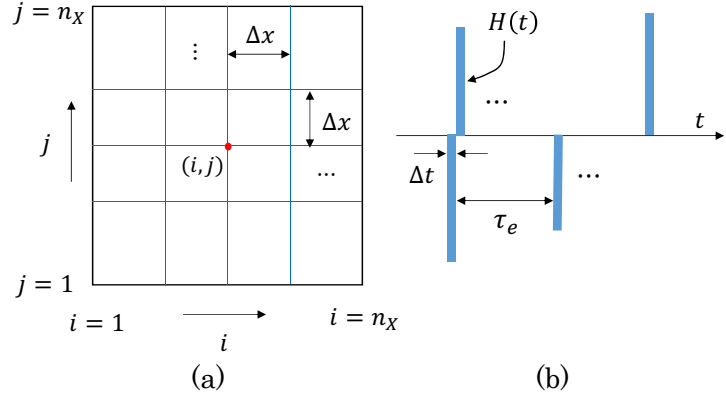}
\caption{ (a) 2D regular square lattice of size $N\!=\!n_X\times n_X$, where a lattice site is represented by $(i,j)$ and the lattice spacing is ${\it \Delta}x$ in both the $i$ and $j$ directions. (b) The stochastic variable $H(t)$ is understood to have an expectation value of zero; however, its square is finite (Eq. (\ref{Brownian-motion})). The force $\eta(t)$ randomly fluctuates inside a narrow square of width ${\it \Delta} t$ in (b), and its time integral $H(t)$ can be intuitively understood as an impulse. The macroscopic relaxation time $\tau_e$ is considerably longer than the width ${\it \Delta} t$. } 
 \label{fig-4}
\end{center}
\end{figure}

To obtain the discrete form of the NS equation in Eq. (\ref{NS-eq}) on a 2D regular square lattice (Fig. \ref{fig-4}(a)), we introduce the quantity
\begin{eqnarray}
\label{Integral-GRF}
\vec H_{i,j}(t)=\int_{t}^{t+{\it \Delta} t}\vec \eta_{i,j}(t) dt,
\end{eqnarray}
where $\eta_{i,j}(t)$ denotes the random force on the fluid particle at lattice site $({i,j})$ at time $t$ (Fig. \ref{fig-4}(b)). Note that since the representation of the lattice site is changed from $i$ to $(i,j)$, $\vec \eta_i\to \vec \eta_{i,j}$ accordingly.
This $\vec H_{i,j}(t)$ is still considered a stochastic variable. From the expression in Eq. (\ref{Integral-GRF}) and the relation in Eq. (\ref{Gaussian-random-force}), it is easy to obtain
\begin{eqnarray}
\label{Brownian-motion}
\langle {\vec H}_{i,j}(t)\rangle=0, \quad \langle H_{i,j}^2(t)\rangle\!=\!2D{\it \Delta} t,
\end{eqnarray}
which are typical characteristics of stochastic variables. The first relation comes from the fact that ${\vec \eta}_{ij}$ corresponds to Gaussian white noise with a mean value of zero, and the time integral and the expectation value operation $\langle \cdots \rangle$ are assumed to be commutative.
If we rewrite $H_{i,j}(t)$ in Eq. (\ref{Integral-GRF}) as
\begin{eqnarray}
\label{Integral-GRF-discrete}
 \vec H_{i,j}(t)=\vec \eta_{i,j}(t){\it \Delta} t
\end{eqnarray}
and substitute this $H_{i,j}(t)$ into the second expression in Eq. (\ref{Brownian-motion}), we obtain a finite value
\begin{eqnarray}
\label{discrete-random-force}
\left|\eta_{i,j}^\mu(t)\right|=\sqrt{2D/{\it \Delta} t} 
\end{eqnarray}
for the random Brownian force. This expression is used in the discrete version of Eq. (\ref{NS-eq}), which is given by
\begin{eqnarray}
\label{discrete-NS-eq}
\begin{split}
&\omega_{i,j}(t+{\it \Delta} t)\leftarrow \omega_{i,j}(t) -{\it \Delta} t\left( \vec V\cdot \nabla\right) \omega_{i,j}(t) \\
&+ \nu{\it \Delta} t {\it \Delta} \omega_{i,j}(t) + \sqrt{2D{\it \Delta} t}(\nabla \times \vec g_{i,j}(t))_z, \\
&\omega_{i,j}=-{\it \Delta} \psi_{i,j},
\end{split}
\end{eqnarray}
where $\vec g_{i,j}(t)\!=\!(g_{i,j}^x(t),g_{i,j}^y(t),0)$ and the components of $\vec g_{i,j}$ are given by Gaussian random numbers with mean $0$ and variance $1$. This $\vec g_{i,j}(t)$ is related to $\vec\eta_{i,j}$ as follows:
\begin{eqnarray}
\label{Brownian-force}
\vec\eta_{i,j}{\it \Delta} t=\sqrt{2D{\it \Delta} t} \vec g_{i,j}(t). 
\end{eqnarray}

On the RHS of the first expression in Eq. (\ref{discrete-NS-eq}), the spatial discretization of the second term with respect to the lattice spacing ${\it \Delta} x$ is given by
\begin{eqnarray}
\label{discrete-second-term}
\begin{split}
& -{\it \Delta} t\left(\vec V \cdot\nabla \right) \omega=-{\it \Delta} t\left(\frac{\partial \psi}{\partial y}\frac{\partial \omega}{\partial x}-\frac{\partial \psi}{\partial x}\frac{\partial \omega}{\partial y}\right)\\
&\to -\frac{{\it \Delta} t}{4({\it \Delta} x)^2} \left[\left(\psi_{i,j+1}-\psi_{i,j-1}\right)\left(\omega_{i+1,j}-\omega_{i-1,j}\right) \right. \\
&\left. -\left(\psi_{i+1,j}-\psi_{i-1,j}\right)\left(\omega_{i,j+1}-\omega_{i,j-1}\right) \right]. 
\end{split}
\end{eqnarray}
This term makes almost no contribution to the flow because the velocity is low (no higher than $\sim 100~{\rm \mu m/s}$) in the case of protoplasmic streaming. Thus, the results are expected to be independent of this term, although we include it in the equation for our simulations. The discrete form of the Laplace operator ${\it \Delta}$ acting on $\omega_{ij}$ is given by
\begin{eqnarray}
\label{discrete-3rd-term}
\begin{split}
{\it \Delta}\omega_{i,j}\to (1/{\it \Delta} x)^2 & \left( \omega_{i+1,j} +\omega_{i-1,j} \right.\\
&\left.+\omega_{i,j+1}+\omega_{i,j-1}-4\omega_{i,j}\right), 
\end{split}
\end{eqnarray}
and ${\it \Delta} \psi_{i,j}$ in the second expression in Eq. (\ref{discrete-NS-eq}) has almost the same discrete form. The discrete form of the final term is given by
\begin{eqnarray}
\label{discrete-final-term}
\begin{split}
&(\nabla \times \vec g_{i,j}(t))_z\to \\
&(g_{i+1,j}^y-g_{i-1,j}^y-g_{i,j+1}^x+g_{i,j-1}^x)/(2{\it \Delta} x). 
\end{split}
\end{eqnarray}

Notably, $\sqrt{2D{\it \Delta} t}$ in Eq. (\ref{discrete-NS-eq}) effectively corresponds to the deviation of the random Brownian force. As determined through dimensional analysis, $(\sqrt{2D {\it \Delta}t})^2{\it \Delta}t\!=\!2D({\it \Delta}t)^2$ is the diffusion constant $D_{\rm dif}$ related to the temperature $T$ by means of the Einstein-Stokes-Sutherland formula $D_{\rm dif}\!=\!{k_BT}/{6\pi \mu a}$, which is identified with $2D({\it \Delta}t)^2$. Here, we introduce the notion of the macroscopic relaxation time $\tau_{e}$, which is the time required for the fluid to equilibrate from the resting state to a stationary state compatible with the boundary condition given by the velocity $V_B$ (Fig. \ref{fig-3}(c)), and this $\tau_{e}$ is independent of whether the initial state is the resting state or a random state \cite{Coffey-Kalmykov-CP1993,Feldmanetal-Wiley2006,KArkhipov-JNCS2002,Zaochik-ETFS1997}. According to this definition, $\tau_e$ is proportional to the area (or volume, more generally), in sharp contrast to the standard relaxation time, which is the mean time required for a molecule to return to its original position from a disturbed position.
We replace ${\it \Delta}t$ with $\tau_{e}$ because ${\it \Delta}t$ is a numerically introduced quantity; thus, we have
\begin{eqnarray}
\label{ESS-formula}
2D\tau_{e}^2(=D_{\rm dif})=\frac{k_BT}{6\pi \mu a}, 
\end{eqnarray}
where $\mu(=\! \rho\nu)$ is the viscosity, $a$ is the size of a fluid particle or a group of particles in the fluid, and $k_{B}$ is the Boltzmann constant. Note that $a$ is larger than the size of a molecule such as water because it is obtained by assuming $\tau_{e}$, which is not a microscopic quantity.
The actual value of $D$ assumed in the simulation and its relation to the $D_{\rm dif}$ value reported in Ref. \cite{Goldstein-etal-PRL2008} will be discussed in the results section.

The second equation of Eq. (\ref{discrete-NS-eq}), which is Poisson's equation, is numerically solved by the convergent configuration of the iterations such that
\begin{eqnarray}
\label{SOR}
\begin{split}
&\psi_{i,j}^{(\ell+1)}(t)\leftarrow  \psi_{i,j}^{(\ell)}(t)\\
&+A\left[\frac{\psi_{i+1,j}^{(\ell)}(t)+\psi_{i-1,j}^{(\ell)}(t)+\psi_{i,j+1}^{(\ell)}(t)+\psi_{i,j-1}^{(\ell)}(t)+({\it \Delta }x)^2\omega_{i,j}(t)}{4}-\psi_{i,j}^{(\ell)}(t)\right],
\end{split}
\end{eqnarray}
where the superscript $\ell$ is an integer denoting fictitious time or the number of iterations and the constant $A$ is the acceleration coefficient fixed to $A\!=\!1$. This technique is generally called the successive-over-relaxation (SOR) technique and is equivalent to the Gauss-Seidel method when $A\!=\!1$. 
The convergence criteria will be discussed in the results section.

\subsection{Boundary conditions\label{boundary-cond}}
\begin{figure}[t]
\begin{center}
\includegraphics[width=9.5cm]{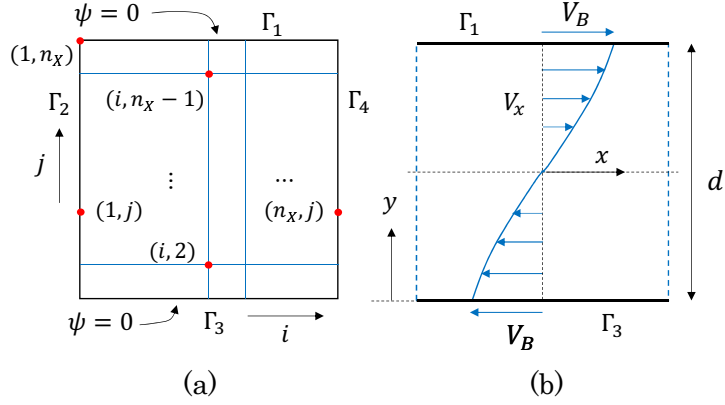}
\caption{  (a) The boundary condition $\psi\!=\!0$ at the boundaries $\Gamma_1$ and $\Gamma_3$. (b) The velocity ${\vec V} \!=\!(V_x,V_y)$ is fixed to ${\vec V} \!=\!(V_B,0)$ at $\Gamma_1$ and to ${\vec V}\!=\!(-V_B,0)$ at $\Gamma_3$. The lattice sites $(i,n_X\!-\!1)$ and $(i,2)$ in (a), close to $\Gamma_1$ and $\Gamma_3$, respectively, are used to enforce the boundary conditions for the variable $\omega$ in Eq. (\ref{boundary-cond-1}). }
 \label{fig-5}
\end{center}
\end{figure}
The boundary conditions for the variables $\omega$ and $\psi$ at the boundaries $\Gamma_1$ and $\Gamma_3$ (Fig.~\ref{fig-5}(a)) are given by
\begin{eqnarray}
\label{boundary-cond-1}
\begin{split}
&\omega_{i,{\rm {n_X}}} =- \frac{2}{({\it \Delta} x)^2} \left( \psi_{i,{\rm {n_X}}-1}+\left| V_B\right| {\it \Delta} x \right), \\
& \qquad \psi_{i,{\rm {n_X}}}=0 \quad ({\rm on}\; \Gamma_1),\\
&\omega_{i,1} =- \frac{2}{({\it \Delta} x)^2} \left( \psi_{i,2}+\left| V_B\right| {\it \Delta} x \right), \quad \psi_{i,1}=0 \quad ({\rm on}\; \Gamma_3),
\end{split}
\end{eqnarray}
where the velocity $\vec V$ at the boundary is given by
\begin{eqnarray}
\label{boundary-velocity}
\vec V=(V_B,0) \quad {\rm on}\; \Gamma_1, \quad \vec V=(-V_B,0) \quad {\rm on}\; \Gamma_3. 
\end{eqnarray}
Note that the third component of $\vec V$ is henceforth assumed to be zero and is thus neglected in all expressions for simplicity.

The reason why the stream function $\psi$ can be fixed to $\psi\!=\!0$ on $\Gamma_1$ and $\Gamma_3$ in Eq. (\ref{boundary-cond-1}) is as follows: The function $\psi$ is not uniquely fixed in the domain because the velocity is given by the first-order differentials in Eq. (\ref{stream-function}). For this reason, $\psi$ and $\psi\!+\!f_0$ are exactly equivalent for any constant $f_0$ in the sense that both $\psi$ and $\psi\!+\!f_0$ correspond to the same velocity configuration. Therefore, if $\psi_{1, n_X}$ is nonzero such that $\psi_{1, n_X}\!=\!c_0$ at $(1,n_X)\in \Gamma_1$, then $\psi$ can be replaced by $\psi+f_0$ with $f_0\!=\!-c_0$, and hence, we have $\psi_{1, n_X}\!=\!0$ (Fig. \ref{fig-5}(a)). It is also easy to check that $\psi_{i, n_X} \!=\!0$ $(i>1)$ because of the boundary condition $V_y\!=\!0$ on $\Gamma_1$ in Eq. (\ref{boundary-velocity}). On $\Gamma_3$, we also have $\psi_{i,1}\!=\!0$ because of the symmetry argument under a rotation by $\pi$ around the $z$-axis perpendicular to the domain. Evidently, no gravitational force is considered, and no asymmetry is expected in the random Brownian force; therefore, this rotational symmetry of $\psi$ is naturally expected.

Note that the expressions for $\omega$ in Eq. (\ref{boundary-cond-1}) are well known and that the expression for $\omega_{i,{\rm {n_X}}}$ on the boundary $\Gamma_1$ is obtained by means of Taylor expansion and the second expression in Eq. (\ref{NS-eq}) as follows:
\begin{eqnarray}
\label{Taylor-exp}
\begin{split}
&\psi(x,y-{\it \Delta} y)\\
&=\psi(x,y)-\frac{\partial \psi}{\partial y}(x,y){\it \Delta} y + \frac{1}{2} \frac{\partial^2 \psi}{\partial y^2}(x,y)\left({\it \Delta} y \right)^2+\cdots \\
&=0+V_B {\it \Delta} x -  \frac{1}{2} \omega(x,y)\left({\it \Delta} x \right)^2+\cdots, \quad {\rm for}\; (x,y)\in \Gamma_1,
\end{split}
\end{eqnarray}
where ${\it \Delta} y\!=\!{\it \Delta} x$ is assumed and $V_x\!=\!- {\partial \psi}/{\partial y}$ in Eq. (\ref{stream-function}) is used. The expression for $\omega_{i,1}$ on $\Gamma_3$ in Eq. (\ref{boundary-cond-1}) can be obtained in the same manner.

On the boundaries $\Gamma_2$ and $\Gamma_4$, periodic boundary conditions along the horizontal or $i$ direction are assumed, such that
\begin{eqnarray}
\label{periodic-boundary-cond}
\begin{split}
&\omega_{{\rm {n_X}}+1,j}=\omega_{1, j}, \quad  \psi_{{\rm {n_X}}+1,j}=\psi_{1, j},\\
&\omega_{-1, j}=\omega_{{\rm {n_X}},j}, \quad  \psi_{-1, j}=\psi_{{\rm {n_X}},j}.\\
\end{split}
\end{eqnarray}
These conditions imply that the lattice sites $(1,j)$ on $\Gamma_2$ and $({\rm {n_X}},j)$ on $\Gamma_4$ are adjacent to each other (Fig. \ref{fig-5}(a)).

\subsection{Physical and simulation units   \label{phys-sim-units}}
In the actual process of protoplasmic streaming, length and time are measured in units of ${\rm m}$ and ${\rm s}$, respectively, while the corresponding values in the simulations are ${\rm \alpha~m}$ and ${\rm \beta~s}$, where $\alpha$ and $\beta$ are positive numbers. We use ${\rm \alpha m}$ and ${\rm \beta s}$ to denote the corresponding so-called simulation units. The transformation rules from ${\rm m}$ and ${\rm  s}$ to ${\rm \alpha m}$ and ${\rm \beta s}$ are given as follows:
\begin{eqnarray}
\label{simu-units}
1~{\rm m}=\alpha^{-1}~{\rm \alpha m},\quad 1~{\rm s}=\beta^{-1}~{\rm \beta s}. 
\end{eqnarray}

The numerical results should be independent of the values of $\alpha$ and $\beta$, which will be discussed in greater detail later. We consider the physical parameters
\begin{eqnarray}
\label{parameters}
\begin{split}
&\nu~[{\rm m^2}/{\rm s}], \quad V_B~[{\rm m}/{\rm s}], \quad  d~[{\rm m}], \quad \tau~[{\rm s}], \\
&D~[{\rm m^2}/{\rm s^3}],\quad   {\it \Delta} x~[{\rm m}], \quad {\it \Delta} t~[{\rm s}]. 
\end{split}
\end{eqnarray}
The kinematic viscosity coefficient $\nu$ is explicitly included in Eq. (\ref{discrete-NS-eq}). The velocity $V_B$ is necessary in the boundary conditions illustrated in Figs. \ref{fig-3}(c) and \ref{fig-5}(b). The third parameter, $d$, is the diameter illustrated in Fig. \ref{fig-5}(b).
The fourth parameter, $\tau$, is set equal to the macroscopic relaxation time $\tau_{e}$ introduced in the final part of Section \ref{discretization} and numerically corresponds to the equilibration time of the fluid (see Fig. \ref{fig-4}(b)).
This length of time should be no less than $\tau_{e}$; otherwise, no equilibrium will be achieved in the numerical simulations. $D$ in Eq. (\ref{ESS-formula}) must be expressed in units of ${\rm m^2}/{\rm s^{3}}$ because each term in Eq. (\ref{discrete-NS-eq}) must have units of ${1}/{\rm s}$, which is clear from the fact that $\omega$ is expressed in units of ${1}/{\rm s}$.

The final two parameters in Eq. (\ref{parameters}), ${\it \Delta} x$ and ${\it \Delta} t$, are necessary only in the simulations; however, these parameters do have indirect counterparts in the actual experimental phenomena. Specifically, these parameters have the following relations to the physical parameters $d$ and $\tau$:
\begin{eqnarray}
\label{sim-parameters}
 {\it \Delta} x= \frac{d}{n_X}, \quad {\it \Delta} t=\frac{\tau}{n_T}, 
\end{eqnarray}
where $n_X$ is the total number of lattice points on one edge of the lattice and $n_T$ is the total number of iterations. The parameter $\tau$ is simply related to the convergence of the time evolution corresponding to a set of random Brownian forces $\{ \eta_{ij}(t)\}$, and it physically corresponds to the macroscopic relaxation time, which represents the typical time scale of the phenomenon, as mentioned above. This convergence is controlled by the small parameter $\varepsilon$, which will be discussed later in the results section. Therefore, exact information on $\tau$ or the macroscopic relaxation time $\tau_{e}$ is unnecessary, at least in simulations. Indeed, as long as $\varepsilon$ is sufficiently small, then a configuration that is randomized by Brownian forces can correctly converge or reach equilibrium. Moreover, if ${\it \Delta} t$ ($n_T$) is excessively large (small), then the simulation will not converge; hence, ${\it \Delta} t_0$ should be fixed to ${\it \Delta} t_0\!\leq\!{\it \Delta} t_{\rm cr}$. This ${\it \Delta} t_{\rm cr}$ is considered the maximal time step satisfying the convergence criterion. Thus, the temporal discretization or time evolution is subtle compared to the spatial discretization. Both of the discretizations will be discussed in greater detail in the results section.

Here, we introduce the symbol $E$ to represent the experimental parameters and the symbol $S$ to represent the simulation parameters, such that
\begin{eqnarray}
\label{X-parameter}
\begin{split}
&E=(\nu,  V,  d), \\
&S=(\nu,  V, D, {\it \Delta} x, {\it \Delta} t).
\end{split}
\end{eqnarray}
In principle, the parameter $D$ should be included in $E$; however, it is unknown for the flows under study. Therefore, $D$ is instead included in $S$ as an input for the simulations.

\subsection{Invariance under unit transformations \label{invariance}}
To discuss the invariance of the simulation results obtained using the discrete NS equation in Eq. (\ref{discrete-NS-eq}) under unit transformations, we use the notion of scale transformations for the parameters ${\rm m}$, ${\rm s}$, $n_X$, and $n_T$  introduced in Sect. \ref{phys-sim-units}. These scale transformations are defined in terms of positive numbers $\alpha$, $\beta$, $\gamma$, and $\delta (>0)$ such that
\begin{eqnarray}
\label{scale-changes}
{\rm m, s}, n_X, n_T \to {\rm \alpha m, \beta s}, \gamma n_X, \delta n_T.
\end{eqnarray}
The first two transformations, (${\rm m, s})\!\to\!({\rm \alpha m, \beta s}$), represent changes to the units of length and time. The third transformation, $n_X\!\to\!\gamma n_X$, represents a change in lattice size, which corresponds to a change in the lattice spacing ${\it \Delta} x\!\to\!\gamma^{-1}{\it \Delta} x$ in Eq. (\ref{sim-parameters}). The final transformation, $n_T\!\to\!\delta n_T$, represents a change in the time step ${\it \Delta} t\!\to\!\delta^{-1}{\it \Delta} t$ by means of Eq. (\ref{sim-parameters}).
$\gamma$ and $\delta$ for $n_X$ and $n_T$ are necessary in addition to $\alpha$ and $\beta$ for ${\rm m}$ and ${\rm s}$ because the simulation results should be independent of the lattice spacing ${\it \Delta} x$ and the time step ${\it \Delta} t$. For a change in ${\it \Delta} x$, for example, the scaling of $n_X$ by $\alpha$ can be used because of the relation ${\it \Delta} x~[{\rm m}]\!=\!d/n_X~[{\rm m}]\!=\!\alpha^{-1}d/n_X~[{\rm \alpha m}]\!=\!d/(\alpha n_X)~[{\rm \alpha m}]$. Indeed, this relation implies that a unit transformation by $\alpha$ can be understood as a change in the lattice size $n_X$. However, in this case, it is impossible to observe the dependence of the results on the lattice size $n_X$ without affecting other parameters that also depend on the length unit, which is why $\gamma$ is necessary in addition to $\alpha$.
Moreover, as will be shown below in further detail, the rescaling of ${\it \Delta} x$ always affects the force strength $D$, which implies that we need to modify $D$ as well as ${\it \Delta} x$ to observe the dependence of the results on ${\it \Delta} x$. This behavior arises from the fact that the order of ${\it \Delta} x$ in the Brownian force term is different from that in the other terms; this difference is simply due to the difference in the order of spatial differentials such as ${\it \Delta}$ and $\nabla$.
The dependence of the results on ${\it \Delta} t$ suffers from the same problem, as will also be studied in detail later. In this case, the difference in the order of ${\it \Delta} t$ originates from the fact that the Brownian force is represented by a stochastic variable, as described in Section \ref{discretization}.

First, we rewrite Eq. (\ref{discrete-NS-eq}) by explicitly including the lattice spacing ${\it \Delta} x$, such that
\begin{eqnarray}
\label{discrete-NS-right}
\begin{split}
&\omega_{i,j}\leftarrow\omega_{i,j}+\frac{{\it \Delta} t }{({\it \Delta} x)^2} \left( \psi\cdots\right)\left( \omega\cdots\right)+\\
& \nu\frac{{\it \Delta} t }{({\it \Delta} x)^2} \left( \omega\cdots\right) + \frac{\sqrt{2D{\it \Delta} t}}{{\it \Delta} x}\left( g\cdots\right)\; [1/{\rm s}],
\end{split}
\end{eqnarray}
where $({\it \Delta} x)^{-2}\left( \psi\cdots\right)\left( \omega\cdots\right)$ on the RHS represents the spatial discretization in Eq. (\ref{discrete-second-term}) and $({\it \Delta} x)^{-2}\left( \omega\cdots\right)$ and $({\it \Delta} x)^{-1}\left( g_z\cdots\right)$ represent the discretizations of the Laplacian ${\it \Delta}\omega$ and the rotation $\left(\nabla \times {\vec g}\right)_z$, respectively. The symbol $[1/{\rm s}]$ represents the overall unit of the terms, which, in Eq. (\ref{discrete-NS-right}), are written with the physical units $[{\rm m}]$ and $[{\rm s}]$.

Under the scale transformations in Eq. (\ref{scale-changes}), ${\it \Delta} x(=\!d/n_X)~[{\rm m}]$ and ${\it \Delta} t(=\!\tau/n_T)~[{\rm s}]$ are replaced by $\alpha^{-1}\gamma^{-1}{\it \Delta} x~[{\rm \alpha m}]$ and $\beta^{-1}\delta^{-1}{\it \Delta} t~[{\rm \beta s}]$, and we also have $\nu~[{\rm m^2}/{\rm s}]\!=\!\alpha^{-2}\beta \nu~[{\rm (\alpha m)^2}/{\rm\beta s}]$ and $D~[{\rm m^2}/{\rm s^3}]\!=\!\alpha^{-2}\beta^3D~[{\rm (\alpha m)^2}/{\rm( \beta s)^3}]$. The units of $\psi$ and $\omega$ are ${\rm m^2}/{\rm s}$ and ${1}/{\rm s}$, respectively; therefore, we have $\psi~[{\rm m^2}/{\rm s}]\!=\!\alpha^{-2} \beta\psi~[{\rm (\alpha m)^2}/{\rm \beta s}]$ and $\omega~[{1}/{\rm s}]\!=\!\beta\omega~[{1}/{\rm \beta s}]$.
From these expressions and Eq. (\ref{discrete-NS-right}), we obtain
\begin{eqnarray}
\label{discrete-NS-right-scaled}
\begin{split}
&\omega_{i,j}\leftarrow\omega_{i,j} +\gamma^2\delta^{-1}\frac{{\it \Delta} t }{({\it \Delta} x)^2} \left( \psi\cdots\right)\left( \omega\cdots\right)  \\
&+ \gamma^2\delta^{-1}\nu\frac{{\it \Delta} t }{({\it \Delta} x)^2} \left( \omega\cdots\right) +\frac{\sqrt{2\gamma^2\delta^{-1}D{\it \Delta} t}}{{\it \Delta} x}\left( g\cdots\right)  \; [{1}/{\rm \beta s}],
\end{split}
\end{eqnarray}
where the common factor $\beta$ is eliminated from both sides. In the case of $\gamma\!=\!1$ and $\delta\!=\!1$, nothing is changed except that the units are changed from ${\rm m}$ and ${\rm s}$ to ${\rm\alpha m}$ and ${\rm\beta s}$. The problem is the case of $\gamma\!\not=\!1$ or $\delta\!\not=\!1$, where the factor $\sqrt{\gamma^2\delta^{-1}}$ in the final term is different from the factor $\gamma^2\delta^{-1}$ in the second and third terms. However, if $D$ transforms as $D\to\gamma^2\delta^{-1}D$ under the scale transformation $(n_X,  n_T)\to(\gamma n_X, \delta n_T)$, then we obtain a common factor of $\gamma^2\delta^{-1}$ in all three of these terms on the RHS of Eq. (\ref{discrete-NS-right-scaled}). In this case, we have
\begin{eqnarray}
\label{discrete-NS-right-scaled-factor}
\begin{split}
&\omega_{i,j}\leftarrow\omega_{i,j} +\gamma^2\delta^{-1}\frac{{\it \Delta} t }{({\it \Delta} x)^2} \left( \psi\cdots\right)\left( \omega\cdots\right)  \\
&+ \gamma^2\delta^{-1}\nu\frac{{\it \Delta} t }{({\it \Delta} x)^2} \left( \omega\cdots\right) +\gamma^2\delta^{-1}\frac{\sqrt{2D{\it \Delta} t}}{{\it \Delta} x}\left( g\cdots\right)  \; [{1}/{\rm \beta s}];
\end{split}
\end{eqnarray}
therefore, the convergent numerical solution is expected to remain unchanged. Indeed, in such a stationary or equilibrium configuration, the term $\omega_{i,j}(t\!+\!{\it \Delta} t)$ on the LHS is expected to be identical to the first term $\omega_{i,j}(t)$ on the RHS; hence, the common factor $\gamma^2\delta^{-1}$ in the remaining terms can be dropped.

The velocity $V$ for the boundary conditions and the diameter $d$ are included in the parameters $E$ or $S$ in Eq. (\ref{X-parameter}), and their scaling properties under unit transformation are given by $V~[{\rm m}/{\rm s}]\!=\!\alpha^{-1}\beta V~[{\rm \alpha m}/{\rm \beta s}]$ and $d~[{\rm m}]\!=\!\alpha^{-1}d~[{\rm \alpha m}]$.
Thus, under the scale transformations in Eq. (\ref{scale-changes}), the RHS of Eq. (\ref{discrete-NS-eq}) remains unchanged in the equilibrium configuration if the parameters $S=(\nu,  V,   D, {\it \Delta} x, {\it \Delta} t)$ scale as follows:
\begin{eqnarray}
\label{param-changes}
\begin{split}
&(\nu,  V,  D, {\it \Delta} x, {\it \Delta} t) \to \\
&(\alpha^{-2}\beta\nu,  \alpha^{-1}\beta V,   \alpha^{-2}\beta^3\gamma^2\delta^{-1} D, \alpha^{-1}\gamma^{-1}{\it \Delta} x,\\
& \beta^{-1}\delta^{-1}{\it \Delta} t).
\end{split}
\end{eqnarray}

Next, we introduce the notion of equivalence in the simulation data. The simulation data obtained by solving Eq. (\ref{discrete-NS-eq}) are denoted by $(\omega,\psi)$, while the experimental velocity data are denoted by ${\rm Exp}(E)$ because the experimental data ${\rm Exp}$ are characterized by the parameters $E$ in Eq. (\ref{X-parameter}).
We define the term {\it equivalent} as follows:
\\
Two sets of simulation data $(\omega_1,\psi_1)$ and $(\omega_2,\psi_2)$ are {\it equivalent} if the following conditions are satisfied:\\
(i) The histogram of the normalized velocity $V_x$ distribution and\\
(ii) the dependence of the normalized $V_x$ on $y$ \\
for $(\omega_1,\psi_1)$ are identical to those for $(\omega_2,\psi_2)$.

Thus, we have proven the following statement:

\noindent
\begin{enumerate}
\item[(A)]
The solution $(\omega,\psi)$ of Eq. (\ref{discrete-NS-eq}) remains unchanged if and only if the parameters $S=(\nu,  V,  D, {\it \Delta} x, {\it \Delta} t)$ scale in accordance with Eq. (\ref{param-changes}) under the scale transformations in Eq. (\ref{scale-changes}).
\end{enumerate}

Here, we comment on the dependence of the simulation results on ${\it \Delta} x$, as mentioned above. The parameter $\gamma$ for rescaling ${\it \Delta} x$ was introduced in Sect. \ref{phys-sim-units} to elucidate this dependence. However, we find from statement (A) that ${\it \Delta} x$ alone cannot be changed without affecting the results. The dependence of the results on ${\it \Delta} t$ shows the same behavior as the dependence on ${\it \Delta} x$. This nonstandard situation in regard to ${\it \Delta} t$ arises from the discrete form of the random Brownian force in Eq. (\ref{discrete-NS-eq}) and is typical of such a discrete Langevin equation \cite{Lemos-AmJP1997,Wilhelm-Springer1989,Metzler-PhysRep2000,KGWilson-PRD1985,Ukawa-Fukugita-PRL1985,Hofler-Schwarzer-PRE2000,Koibuchi-JPG1987,Nagahiro-PRE2007}.

The experimental data ${\rm Exp}(E)$ can also be grouped into equivalent classes in the same way. Specifically, two different sets of experimental data ${\rm Exp}(E_1)$ and ${\rm Exp}(E_2)$ are considered equivalent if conditions (i) and (ii) are satisfied. 

A similar notion of equivalence can be introduced for the parameters $S$ in Eq. (\ref{X-parameter}). 
Two sets of parameters $S_1=(\nu_1,  V_1,   D_1, {\it \Delta} x_1, {\it \Delta} t_1)$ and $S_2=(\nu_2,  V_2, D_2, {\it \Delta} x_2, {\it \Delta} t_2)$ are called {\it equivalent} if there exists a set of positive numbers $\alpha$, $\beta$, $\gamma$, and $\delta$ (all $>0$) such that
\begin{eqnarray}
\label{param-changes-2}
\begin{split}
&(\nu_1,  V_1,  D_1, {\it \Delta} x_1, {\it \Delta} t_1) \\ = 
&(\alpha^{-2}\beta\nu_2,  \alpha^{-1}\beta V_2,  \alpha^{-2}\beta^3 \gamma^2\delta^{-1}D_2,\\
& \alpha^{-1}\gamma^{-1}{\it \Delta} x_2, \beta^{-1}\delta^{-1}{\it \Delta} t_2).
\end{split}
\end{eqnarray}
This equivalence is denoted by $S_1\equiv S_2$. It is easy to confirm that $S_2\equiv S_1$ if $S_1\equiv S_2$ because $\alpha$ and $\beta$ can be inverted to $\alpha^{-1}$ and $\beta^{-1}$. $S_1\!=\!S_2$ if and only if $\alpha\!=\!\beta\!=\!\gamma\!=\!\delta\!=\!1$.
A set of parameters equivalent to $S=(\nu,  V,  D, {\it \Delta} x, {\it \Delta} t)$ is written as $\bar S=(\bar\nu,  \bar V,   \bar D, \bar{{\it \Delta} x}, \bar{{\it \Delta} t})$.

Notably, two different parameter sets $S_i$ $(i=1,2)$ produce the same solution $(\omega,\psi)$ for the discrete NS equation in Eq. (\ref{discrete-NS-eq}) if these $S_i$ $(i=1,2)$ are equivalent; in other words, two solutions $(\omega_1,\psi_1)$ and $(\omega_2,\psi_2)$ are equivalent if the corresponding parameter sets $S_i$ $(i=1,2)$ are equivalent. In this sense, the solution to Eq. (\ref{discrete-NS-eq}) depends only on the equivalent class of parameters $\bar S$. If one of the parameters is transformed as $\nu\to \nu_1(\not=\nu)$ in $\bar S$, then the parameters $(\bar\nu_1,  \bar V,  \bar d,  \bar D, \bar{{\it \Delta} x}, \bar{{\it \Delta} t})$ are not equivalent to the original $\bar S$.

Finally, in this subsection, we introduce the notation
\begin{eqnarray}
\label{simulatable}
\begin{split}
&{\rm Exp}(E)\simeq S_0 \\
& {\rm for} \quad E=(\nu,  V,  d), \; S_0=(\nu_0,  V_0,  D_0, {\it \Delta} x_0, {\it \Delta} t_0), 
\end{split}
\end{eqnarray}
which means that the experimentally observed data of the velocity distribution are equivalent to the simulation data in the sense defined above. The meaning of the expression ${\rm Exp}(E)\simeq S_0$ is that ``the experimental data ${\rm Exp}(E)$ are successfully simulated with the parameters $S_0$''.

\subsection{Unique solution to the Navier-Stokes equation    \label{uniqueness}}
The problem that we would like to clarify is how many parameters are sufficient to simulate the real experimental data ${\rm Exp}(E)$. We must consider this problem because the solution to Eq. (\ref{discrete-NS-eq}) depends on many parameters $S\!=\!(\nu,  V,   D, {\it \Delta} x, {\it \Delta} t)$, even though only their equivalent classes are meaningful. One possible answer is that only the parameter $D$ must be varied to enable the simulation of arbitrary ${\rm Exp}(E_e)$ data, while the remaining parameters can be fixed to the parameters $S_0$ used to simulate certain existing experimental data ${\rm Exp}(E_{e,0})$, which are not always identical to ${\rm Exp}(E_e)$. This process is described in more detail in the following statement.

\noindent
\begin{enumerate}
\item[(B)]
Let $E_{e,0}=(\nu_{e,0},  V_{e,0},  d_{e,0})$ be a set of parameters that characterize ${\rm Exp}(E_{e,0})$, and let $S_0$ be a set of parameters given by $S_0=(\nu_0,  V_0,   D_0, {\it \Delta} x_0, {\it \Delta} t_0)$. In this situation, if ${\rm Exp}(E_{e,0})\!\simeq\! S_0$, then for any experimental data ${\rm Exp}(E_e)$, with $E_e\!=\!(\nu_e,  V_e,  d_e)$ and $D_e$, and for any given set of $(n_X, n_T)$, there exists a unique $D_{\rm sim}$ such that ${\rm Exp}(E_e)\simeq (\nu_0,   V_0,    D_{\rm sim}, {{\it \Delta} x_0}, {{\it \Delta} t_0})$.
\end{enumerate}

Statement (B) indicates that only one parameter, $D_{\rm sim}$, must be varied to simulate arbitrary experimental data ${\rm Exp}(E_e)$. Proof of statement (B) and further details are provided in Appendix \ref{append-A}.

\section{Simulation results\label{results}}
\subsection{Computational procedure and Langevin simulation technique\label{Langevin-simulation}}
The simulation program is written in Fortran, and the Gaussian random numbers are generated by a Box-Muller transformation of uniform random numbers \cite{Fortran-PC}.  
The flow field is exactly the same as that for Couette flow if the Brownian force $\vec \eta$ is zero, and whether the results obtained under the condition $\vec \eta\!=\!\vec 0$ are compatible with the expected solution will be determined below. In the case of $\vec \eta\!\not=\!\vec 0$, the validity of the technique depends on whether the discrete Langevin dynamics with the term in Eq. (\ref{Brownian-force}) are meaningful. This problem has been shown to be meaningful in particle physics in Refs. \cite{Lemos-AmJP1997,Wilhelm-Springer1989,Metzler-PhysRep2000,KGWilson-PRD1985,Ukawa-Fukugita-PRL1985,Hofler-Schwarzer-PRE2000, Koibuchi-JPG1987,Nagahiro-PRE2007}.

The convergent configuration of the variable $\{\omega\}$ for the first equation in Eq. (\ref{discrete-NS-eq}) is obtained using the small number
\begin{eqnarray}
\label{small-number}
 \varepsilon=1\times 10^{-9}
\end{eqnarray}
for a time step of the NS equation such that
\begin{eqnarray}
\label{conv-time-step}
 \left(1/N\right)\sum_{ij}\left|1-\frac{\omega_{ij}(t+{\it \Delta}t)}{\omega_{ij}(t)}\right|<\varepsilon,
\end{eqnarray}
for $\omega$, and the same small number $\varepsilon$ is also assumed for the variable $\psi$ such that
\begin{eqnarray}
\label{conv-for-psi}
\left(1/N\right)\sum_{ij}\left|1-\frac{\psi_{ij}(t+{\it \Delta}t)}{\psi_{ij}(t)}\right|< \varepsilon,
\end{eqnarray}
from which the convergent configuration $\{\psi\}$ is obtained. $N(=\!\sum_{ij}1\!=\!n_X \!\times\! n_X)$ in Eqs. (\ref{conv-time-step}) and (\ref{conv-for-psi}) represents the total number of vertices or the size of the lattice, and the subscript $ij$ denotes a particular lattice site.
 The convergence of the SOR technique in Eq. (\ref{SOR})  for Poisson's equation, which is the second equation in Eq. (\ref{discrete-NS-eq}), is given by the same number $\varepsilon$ such that 
\begin{eqnarray}
\label{conv-Pois}
\left(1/N\right)\sum_{ij}\left|1-\frac{\psi_{ij}^{(\ell+1)}(t)}{\psi_{ij}^{(\ell)}(t)}\right|< \varepsilon
\end{eqnarray}
at each time step ${\it \Delta}t$.

Note that $\varepsilon$ is related to $\tau$ in Eq. (\ref{parameters}). Indeed, $\varepsilon$ determines the total number of iterations $n_T$, which satisfies $n_T\!=\!\tau / {\it \Delta}t$ and, hence, depends only on $\tau$ for a fixed ${\it \Delta}t$. From this relation, it is clear that $\varepsilon$ depends on $\tau$. If $\varepsilon$ is excessively large, the iterative solution processes for the NS equation and the Poisson equation will not converge. By contrast, if $\varepsilon$ is excessively small, then $n_T$ will be very large, resulting in time-consuming simulations. The number $\varepsilon$ in Eq. (\ref{small-number}) is considered sufficiently small because if it is increased by a factor of ten, i.e., replaced with $10\varepsilon$, the results will remain unchanged. The dependence on the time step ${\it \Delta}t$ will be checked separately in Sec. \ref{time-step-dep}.

The mean value of a physical quantity $Q$ is calculated as
\begin{eqnarray}
\label{mean-value}
 Q  = (1/n)\sum _{k=1}^n Q_k,
\end{eqnarray}
where $Q_k$ is the $k$-th sample corresponding to the $k$-th convergent configuration $\{\omega,\psi\}_k$ and $n$ denotes the total number of samples. The configuration $\{\omega,\psi\}_k$, corresponding to a given set of Gaussian random forces $\{g\}_k$, is obtained by iterating the time step ${\it \Delta}t$ and solving the Poisson equation in Eq. (\ref{discrete-NS-eq}) once for every time increment of ${\it \Delta}t$. Thus, by repeating these two steps, we obtain the sample $Q_k$ in Eq. (\ref{mean-value}) from the convergent configuration $\{\omega,\psi\}_k$. The total number $n$ of samples lies in the range of $1\!\times\!10^4\!\leq\! n\!\leq\! 2\!\times\! 10^4$ for all simulations. Note that the formula given in Eq. (\ref{mean-value}) for the mean value is exactly the same as that for the Monte Carlo simulation technique for statistical mechanical models \cite {Metropolis-JCP-1953,Landau-PRB1976}, and these two techniques are known to be equivalent \cite{Lemos-AmJP1997,Wilhelm-Springer1989,Metzler-PhysRep2000,KGWilson-PRD1985,Ukawa-Fukugita-PRL1985,Hofler-Schwarzer-PRE2000,Koibuchi-JPG1987}. This is the reason why we call Eq. (\ref{NS-eq}) (or Eq. (\ref{discrete-NS-eq})) the stochastic NS equation.

We note that the mean value of the velocity $\vec V$, for example, is independent of the order of the calculations:
\begin{enumerate}
\item[(i)] The mean value $\psi$ may first be calculated from the configurations $\{\omega,\psi\}_k$ $(k\!=\!1,n)$, and $\vec V$ can then be calculated using this mean value configuration $\psi$.
\item[(ii)] $\{{\vec V}\}_k$ may first be calculated from the $k$-th configuration $\{\omega,\psi\}_k$, and $\vec V$ can then be calculated as the mean value of the $\{{\vec V}\}_k$ $(k\!=\!1,n)$.
\end{enumerate}
This is simply because of the commutativity of the calculation of $\vec V$ using Eq. (\ref{stream-function}) and the mean value calculation using Eq. (\ref{mean-value}).

Regarding the relation to the lattice Boltzmann method (LBM), the stochastic NS equation is obtained simply by including the Brownian force term in the NS equation; hence, this definition is slightly different from that of the LBM, which is a technique for describing fluid flow from the viewpoint of particle mechanics \cite{Succi-LBM2001}. Nevertheless, the stochastic NS equation technique can be regarded as a special case of the LBM because all terms, including the random Brownian force term, are understood to represent forces acting on the fluid particles. However, we do not delve into the details of this problem because our interest is simply focused on reproducing the experimentally observed velocity distribution.

Here, we comment on the dependence of the results on the initial configuration of the variables $\{\omega,\psi\}$. On the boundaries $\Gamma_1$ and $\Gamma_3$, the variables are fixed to certain assumed values in accordance with the boundary conditions. Inside the flow region, two possible initial configurations can be assumed for $\{\omega,\psi\}_k$: $\omega\!=\!\psi\!=\!0$ and the convergent configuration $\{\omega,\psi\}_{k-1}$, where $\{\omega,\psi\}_{0}$ is $\omega\!=\!\psi\!=\!0$. The results, including $n_T$, which is the total number of iterations in the sense of the mean values, are independent of these initial configurations, implying that the macroscopic relaxation time $\tau_e$ is also independent of the initial conditions.

\subsection{Normalized velocity distribution\label{velocity-distribution}}
\begin{figure}[h!]
\begin{center}
\includegraphics[width=12cm]{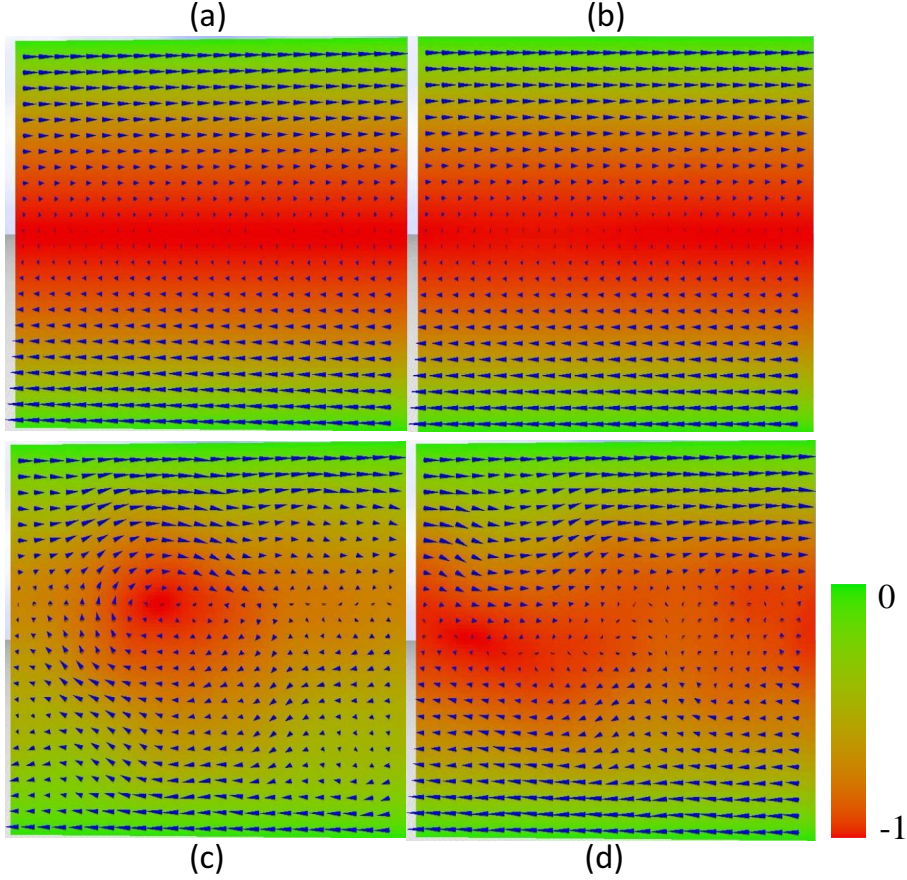}
\caption{ (a) Snapshot of the normalized stream function $\psi$ and the velocity $\vec V$ obtained for $D_{\rm sim}\!=\!0$. (b) The mean values of 1,000 convergent configurations for $D_{\rm sim}\!=\!400$. (c),(d) Snapshots of convergent configurations for $D_{\rm sim}\!=\!400$. Small cones represent $\vec V$, and the colors represent $\psi$.
In the case of $D_{\rm sim}\!=\!0$, no random Brownian force is assumed, and a uniquely determined configuration appears. In contrast, in the case of $D_{\rm sim}\!=\!400$, all of the convergent configurations are slightly different, two examples of which are illustrated by the snapshots in (c) and (d). If $D_{\rm sim}\!=\!400$ is decreased to $D_{\rm sim}\!=\!0$, all of the possible configurations continuously deform to that depicted in (a), while if $D_{\rm sim}\!=\!400$ is continuously increased to $D_{\rm sim}\!=\!4000$, for example, then the vectors in each snapshot become more clearly deformed except in the boundary region; however, no deviation can be observed in the mean values of $\psi$ or the velocity with respect to those depicted in (b) for $D_{\rm sim}\!=\!400$. \label{fig-6}  }
\end{center}
\end{figure}
First, before presenting the velocity distribution, we show snapshots of the flow field in Figs. \ref{fig-6}(a)--(d). Figure \ref{fig-6}(a) shows the snapshot obtained for $D_{\rm sim}\!=\!0$, and Fig. \ref{fig-6}(b) shows the mean $\psi$ values of 1,000 convergent configurations for $D_{\rm sim}\!=\!400$ and the $\vec V$ results calculated using this mean $\psi$. Figures \ref{fig-6}(c) and \ref{fig-6}(d) show snapshots of two different convergent configurations for the same $D_{\rm sim}\!=\!400$. The parameters other than $D_{\rm sim}$ are the same as those used for the velocity distribution, which will be presented below. In the graphics, the flow velocity is represented by small cones, and the stream function is normalized to $-1\!\leq\! \psi\!\leq\!0$ and represented by a gradient between two different colors, with $\psi$ reaching its maximum value of $\psi\!=\!0$ at the boundaries $\Gamma_1$ and $\Gamma_3$. If the random Brownian force is neglected ($\Leftrightarrow D_{\rm sim}\!=\!0$), the flow field is uniquely determined by Eq. (\ref{trivial-sol}). The snapshot in Fig. \ref{fig-6}(a) is compatible with this expected solution. In contrast, for a nonzero $D_{\rm sim}$, the flow field exhibits fluctuations even if it is convergent. The graphic in Fig. \ref{fig-6}(b) appears to be almost the same as that in (a); however, the velocity distributions are different, as will be shown below.
In Figs. \ref{fig-6}(c) and \ref{fig-6}(d), $\vec V$ and $\psi$ are different from those in Fig. \ref{fig-6}(b) and fluctuate from one convergent configuration to another. We emphasize that these fluctuating configurations are ensemble configurations, which are thermally fluctuating in the statistical mechanical sense, and are not always identical to experimentally observed configurations in general. However, they are understood to be some of the possible configurations, and it is interesting that such vortex configurations are included among the ensemble configurations. In some specific cases, a vortex configuration can dominate; however, we do not address this problem in detail. The experimentally observable quantity that we numerically study in this paper is the velocity distribution, which will be presented below.

\begin{figure}[h]
\begin{center}
\includegraphics[width=9.5cm]{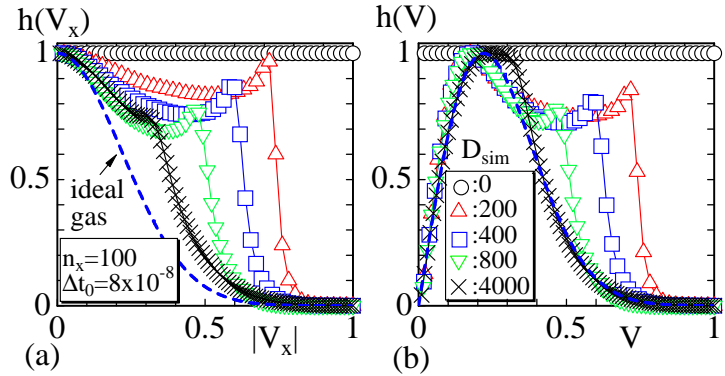}
\caption{ (a) The distribution $h(V_x)$ of $|V_x|$ along the $x$-direction. (b) The distribution $h(V)$ of the length of ${\vec V}$. The lattice size is fixed to $n_X\!=\!100$, and the strength $D_{\rm sim}$ of the random Brownian force is varied from $D_{\rm sim}\!=\!0$ to $D_{\rm sim}\!=\!4000$ in the simulation units. The dashed lines are normalized distributions $\exp(-V_x^2/(2c^2))$ in (a) and $V \exp(-V^2/(2c^2))$ in (b) expected from Maxwell-Boltzmann distribution corresponding to the ideal gas. In these expressions, $c=0.22$ is assumed, and the normalization factor in $V \exp(-V^2/(2c^2))$ is dropped. Due to the violation of the equipartition of energy, the distribution $h(V_x)$ in (a) deviates from the ideal gas behavior even at the largest $D_{\rm sim}\!=\!4000$.}  
 \label{fig-7}
\end{center}
\end{figure}
Now, the main results are presented. In Figs. \ref{fig-7}(a) and \ref{fig-7}(b), we plot the distribution (or normalized histogram) $h(V_x)$ of the absolute velocity $|V_x|$ along the $x$-direction (the longitudinal direction) and the distribution $h(V)$ of the magnitude $V$ of the velocity vector ${\vec V}\!=\!(V_x,V_y)$, respectively.
The discrete expressions are
\begin{eqnarray}
\label{discrete-velocity}
 V_x=-\frac{1}{2{\it \Delta}x} \left( \psi_{i,j+1}-\psi_{i,j-1}\right), \quad
V_y=\frac{1}{2{\it \Delta}x} \left( \psi_{i+1,j}-\psi_{i-1,j}\right),
\end{eqnarray}
which are the standard discrete forms corresponding to the first-order differentials in Eq. (\ref{stream-function}).

The lattice size is fixed to $N\!=\!10000$, with $N\!=\!n_{X}\!\times\! n_{X}$ and $n_{X}\!=\!100$. According to statement (B), the simulation results are expected to depend only on the strength of the random Brownian force, $D_{\rm sim}$; hence, $D_{\rm sim}$ is varied from $D_{\rm sim}\!=\!0$ to $D_{\rm sim}\!=\!4000$ in the simulation units. The lattice size dependence will be presented in Sec. \ref{lat-size-dep}.

The distribution $h(V_x)$ is obtained by constructing a histogram in which the velocity range $0\leq |V_x|\leq v_x^{\rm max}$ is divided into 100 subranges, and the velocity $V_x$ in Eq. (\ref{discrete-velocity}) at lattice point $(i,j)$ is counted in the histogram for every convergent configuration. The maximum velocity $v_x^{\rm max}$ is fixed to $v_x^{\rm max}\!=\!2V_B$, where $V_B$ denotes the velocity at the boundaries $\Gamma_1$ and $\Gamma_3$. The reason for this choice of $v_x^{\rm max}$ is that $v_{x,(i)}^{\rm max}$ in the $i$-th convergent configuration is expected to vary with $i$ instead of remaining constant. The factor of 2 in $2V_B$ for $v_x^{\rm max}$ is sufficiently large because the fluctuations in $v_{x,(i)}^{\rm max}$ are relatively small. Such an assumption regarding the maximum velocity for velocity normalization is unnecessary if the mean value of $\psi$ is calculated first, in accordance with procedure (i) described in the preceding subsection, because the velocity distribution is obtained from a single configuration corresponding to the mean $\psi$ in that case. The histogram $h(V_x)$ is also independent of procedures (i) and (ii) for the calculation of $V_x$.
The height of $h(V_x)$ is normalized such that the maximum height is equal to 1 for each $D_{\rm sim}$, and the horizontal axis $|V_x|$ is similarly normalized using the maximum $|V_x|$ satisfying $h(V_x)\!\not=\!0$ for each $D_{\rm sim}$. The histograms $h(V)$ with respect to $V$ in Fig. \ref{fig-7}(b) are normalized in the same way.

From Fig. \ref{fig-7}(a), we find that $h(V_x)$ is flat for $D_{\rm sim}\!=\!0$; this flat $h(V_x)$ is compatible with the expectation from Eq. (\ref{trivial-sol}). For a sufficiently large $D_{\rm sim}$, the nonflat $h(V_x)$ has a peak only at $V_x\!=\!0$.
However, for intermediate $D_{\rm sim}$ values, another peak appears at $V_x\!\not=\!0$.
If all the ensemble configurations for $D_{\rm sim}\!\not=\!0$ were identical to that in Fig. \ref{fig-6}(a) for $D_{\rm sim}\!=\!0$, then $h(V_x)$ would be expected to be flat. Therefore, configurations including vortices, such as those shown in Figs. \ref{fig-6}(c) and \ref{fig-6}(d), are understood to be the reason for the peaks observed in $h(V_x)$ in numerical studies, even though no such vortex characteristics are apparent in the mean value configuration in Fig. \ref{fig-6}(b). We should note that the flow field shown in Fig. \ref{fig-6}(b) is understood to be slightly different from the observable configurations in experimental measurements of \textit{Nitella} cells due to the imposed simplifications.
The distribution $h(V)$ in Fig. \ref{fig-7}(b) drops to zero, $h(V)\!\to\! 0$ in the limit of $V\!\to\!0$, in contrast to $h(V_x)$ in Fig. \ref{fig-7}(a). This drop is reasonable because the fluid is always moving and there are no fluid particles with zero velocity, $\vec V(t)\!=\!\vec 0$, for all $t$. The drop in $h(V)$ at $V\!\to\!0$ corresponds to the same drop that is visible in the experimentally reported data in Ref. \cite{Mustacich-Ware-PRL1974}. Indeed, with the light scattering technique, not only $\vec V(=\!(V_x,0))$ but also $\vec V(=\!(V_x,V_y))$, which has a small nonzero $V_y$ component, can be detected. 
This drop of $h(V)$ at $V\to 0$ in Fig. \ref{fig-7}(b), for sufficiently large $D_{\rm sim}$, is also consistent with the drop observed in the Maxwell-Boltzmann distribution $V \exp(-V^2/(2c^2))$ expected in the ideal gas. In this expression, the constant $c(=0.22)$ is obtained from the peak position $V$ of $h(V)$ for $D_{\rm sim}\!=\!4000$ in Fig. \ref{fig-7}(b). The simulation result $h(V)$ for $D_{\rm sim}\!=\!4000$ is almost consistent with the dashed line as expected, even though $D_{\rm sim}$ is finite. However, the corresponding $h(V_x)$ considerably deviates from the dashed line $\exp(-V_x^2/(2c^2))$ in Fig. \ref{fig-7}(a). The reason for this deviation is that the equipartition of energy, $\langle V_x^2\rangle\!=\!\langle V_y^2\rangle$, is obviously violated due to the effect of boundary velocity, at least for finite $D_{\rm sim}$. Nevertheless, the ideal gas behavior observed in $h(V)$ for sufficiently large $D_{\rm sim}$ is also reasonable because $D_{\rm sim}$ is proportional to the temperature $T$ as a result of the Einstein-Stokes-Sutherland formula; hence, for sufficiently large $T$, the random Brownian force is expected to be very large compared to the other interactions within the fluid. 
The most important point to note is that the simulation results at finite $D_{\rm sim}(\not=\! 0)$ are located between $\exp(-V_x^2/(2c^2))$ (dashed line) expected from the ideal gas and $h(V_x)\!=\!1 \;(\bigcirc)$ expected from the exact solution of the Couette flow.

We will now discuss the parameters used in the simulations in detail. The viscosity is considered to be almost $100$ times greater than that of water. In Ref. \cite{Kamiya-Kuroda-1973}, the viscosity is reported to be $0.5\leq \mu \leq 1.5~[{\rm dyn \; s /cm^2}]$. Therefore, if the density $\rho$ is assumed to be the same as that of water, then the kinematic viscosity $\nu_{e,0}$ ranges from $\nu_{e,0}\!=\!0.5\!\times\! 10^{-4}~{\rm m^2/s}$ to $\nu_{e,0}\!=\!1.5\!\times\! 10^{-4}~{\rm m^2/s}$. Thus, we assume that $\nu_{e,0}\!=\!1\!\times\! 10^{-4}~{\rm m^2/s}$.
In the experiment conducted by Kamiya \cite{Kamiya-Kuroda-1956}, a cell with a diameter of 0.46 mm was used for the velocity measurements, and 50 ${\rm \mu m}/{\rm s}$ was observed at the boundary. For this reason, a velocity of $V_{e,0}(=\!50~{\rm \mu m}/{\rm s})$ at the boundary and a diameter of $d_{e,0}(\!=500~{\rm \mu m})$ are assumed, as shown in Table \ref{table-2}.
\begin{table}[htb]
\caption{Physical parameters $E_{e,0}\!=\!(\nu_{e,0}, V_{e,0}, d_{e,0})$ assumed in the simulations, expressed in physical units, and the parameters $\alpha_0$ and $\beta_0$ for the conversion to the simulation units.   }
\label{table-2}
\begin{center}
 \begin{tabular}{cccccccc   }
 \hline
 $\nu_{e,0}~[{\rm  m^2}/{\rm  s}]$ & $V_{e,0}~[{\rm \mu m}/{\rm s}]$ & $d_{e,0}~[{\rm \mu m}]$ & $\quad\alpha_0 \quad$ & $\quad\beta_0$   \\
 \hline
 $1\times 10^{-4}$       &  $50$    &   $500$  &  $1\times 10^{-6}$       &  $1\times 10^{-1}$   \\
\hline
\end{tabular}
\end{center}
\end{table}
The parameters $\alpha_0$ and $\beta_0$ in Eq. (\ref{simu-units}) that are used in the simulations are also shown in Table \ref{table-2}. These values, $\alpha_0\!=\!1\!\times\! 10^{-6}$ and $\beta_0\!=\!1\!\times\! 10^{-1}$, imply that the simulation units for length and time are $1~{\rm \mu m}$ and $0.1~{\rm  s}$, respectively. The first three parameters in Table \ref{table-2} are collectively denoted by $E_{e,0}$.
The corresponding $\alpha_0$ and $\beta_0$ values are obtained via Eq. (\ref{alpha-beta-1}) using the parameters $S_0$ shown in the following Table \ref{table-3}.

\begin{table}[htb]
\caption{The parameters $S_0$ used in the simulations;  these values are given in the simulation units. $\gamma_0 n_X$ is written as $n_{X0}$ for simplicity. }
\label{table-3}
\begin{center}
 \begin{tabular}{cccccccccccc   }
 \hline
 $\nu_0~[\frac{\rm (\alpha_0 m)^2}{\rm \beta_0 s}]$ &&\ $V_0~[\frac{\rm \alpha_0 m}{\rm \beta_0 s}]$ &&  ${\it \Delta}x_0~[{\rm \alpha_0 m}]$  &&   ${\it \Delta}t_0~[{\rm \beta_0 s}]$ && $n_{X0}$  \\
 \hline
 $1\times 10^7$       &&  5   &&   5 && $8\times10^{-8}$  && 100 \\
 \hline
\end{tabular}
\end{center}
\end{table}
The lattice spacing ${\it \Delta}x_0~[{\rm \alpha_0 m}]$ is calculated using Eq. (\ref{lattice-spacing-2}). For ${\it \Delta}x_0=5~{\rm \alpha_0 m}$, we have ${\it \Delta}x_{e,0}\!=\!{\it \Delta}x_0\alpha_0^{-1}\!=\!5\times 10^{-6}~{\rm m}$.

There is no need to mention that the parameters $S_0$ in Table \ref{table-3} are also obtained from the parameters $E_{e,0}$ in Table \ref{table-2} and $\alpha_0$ and $\beta_0$. Indeed, it is easy to check that the kinematic viscosity $\nu_0\!=\!1\!\times\! 10^{7}~{\rm \alpha_0^2 m^2/\beta_0 s}$ is obtained from the relation $\nu_{e,0}~[{\rm m^2/s}]=1\!\times\! 10^{-4} \beta_0/\alpha_0^2~[{\rm \alpha_0^2 m^2/\beta_0 s}]=\nu_0~[{\rm \alpha_0^2 m^2/\beta_0 s}]$. The velocity $V_0~[{\rm \alpha_0 m}/{\rm \beta_0 s}]$ and $d_0~[{\rm \alpha_0 m}]$ are also obtained through the relations $V_e\alpha_0^{-1}\beta_0~[{\rm \alpha_0 m}/{\rm \beta_0 s}]\!=\! V_0$ and $d_e\alpha_0^{-1}\!=\!d_0~[{\rm \alpha_0 m}]$, which are not explicitly used in the simulations.

\begin{table}[htb]
\caption{The assumed parameters $D_{\rm sim}$ and $D_{e,0}$, the results $\tau_0$ estimated as $\tau_0\!=\! n_T{\it \Delta}t_0$, and the diameter $a$ of a fluid particle as estimated using Eq. (\ref{ESS-formula}).  }
\label{table-4}
\begin{center}
 \begin{tabular}{cccccccccc   }
 \hline 
$D_{\rm sim}~[\frac{\rm (\alpha_0 m)^2}{({\rm \beta_0 s})^3}]$ && $D_{e,0}~[{\rm m^2}/{\rm s}^3]$ && $n_T{\it \Delta}t_0~[{\rm \beta_0 s}]$ &&   $a~[{\rm m}]$     \\
 \hline
   200 && $2\times10^{-7}$   && $8.7\times10^{-3}$  && $3.6\times10^{-9}$   \\
   400 && $4\times10^{-7}$   && $7.0\times10^{-3}$  && $2.8\times10^{-9}$  \\
   800 && $8\times10^{-7}$   &&  $7.0\times10^{-3}$ && $2.8\times10^{-9}$ \\
 1200 && $1.2\times10^{-6}$ && $1.1\times10^{-2}$  && $7.5\times10^{-10}$ \\
 4000 && $4\times10^{-6}
 $   && $2.6\times10^{-2}$  && $4\times10^{-10}$    \\
\hline
\end{tabular}
\end{center}
\end{table}
Each value of $D_{\rm sim}$ shown in Table \ref{table-4} is fixed as an input to the simulations, and the corresponding physical strength $D_{e,0}$ can be obtained as $D_{e,0}\!=\!D_{\rm sim}\alpha_0^2\beta_0^{-3}$. The parameter $\tau_0$ in Table \ref{table-4} is estimated using the relation $\tau_0\!=\! n_T{\it \Delta}t_0$, where $\tau_0$ is expected to satisfy $\tau_0\leq n_T {\it \Delta}t_{\rm cr}$, as discussed in Section \ref{phys-sim-units}.
Using these $D_{e,0}$ and $\tau_0$ values, we can estimate the diameter $a~[{\rm m}]$ of a fluid particle using Eq. (\ref{ESS-formula}), where the temperature is assumed to be $T\!=\!300~{\rm K}$ and $k_BT\!=\!4.1\times 10^{-21}~{\rm Nm}$. Although the time scale $\tau$ in Eq. (\ref{ESS-formula}) is expected to be smaller than $\tau_0$, we use the results of $\tau_0\!=\! n_T{\it \Delta}t_0$ to calculate $a$.
Nevertheless, the value of $a$ calculated in this way should be larger than the size of a water molecule, i.e., $4\!\times\!  10^{-10}~{\rm m}$, and indeed, we find that almost all the data in Table \ref{table-4} satisfy this condition. 

The diffusion constant $D_{\rm dif}$ in Eq. (\ref{ESS-formula}) can be estimated by assuming that $n_T{\it \Delta}t_0$ is $\tau_e$ as follows: $D_{\rm dif}\!=\!2D_{e,0} (n_T{\it \Delta}t_0 \beta_0) ^2$. For $D_{e,0}\!=\!4\!\times\! 10^7~{\rm m^2/s^3}$ in Table \ref{table-4}, we have $D_{\rm dif}\!\simeq\!4\times 10^{-11}~{\rm m^2/s}\!=\!4\times 10^{-7}~{\rm cm^2/s}$, which is almost 10 times smaller than the estimated value of $D_{\rm dif}\!=\!k_BT/6\pi\mu a\!\simeq\!5.3\times 10^{-6}~{\rm cm^2/s}$ obtained with $\mu\!=\!0.1~{\rm m^2/s}$ and $a\!=\!4\times 10^{-10}~{\rm m}$. Note, however, that this value of $5.3\times 10^{-6}~{\rm cm^2/s}$ is comparable to the value of $10^{-5}~{\rm cm^2/s}$ reported in Ref. \cite{Goldstein-etal-PRL2008}. The deviation between the estimates $D_{\rm dif}\!=\!2D\tau_e^2$ and $D_{\rm dif}\!=\!k_BT/6\pi\mu a$ is expected to shrink in one of the following two possible cases: either $a$ may be considered equal to the radius of a group of water molecules and, hence, should be larger than the radius of a single water molecule, or $\tau_e$ may be slightly larger than $n_T{\it \Delta}t_0 \beta_0$. If either of these conditions is satisfied, then the two estimates are almost compatible, and the notion of $\tau_e$ as adopted in Eq. (\ref{ESS-formula}) is reasonable. Clearly, the first condition, at least, is quite reasonable.

Now, let us comment on the Schmidt number ${\rm S_c}$, which is calculated as the ratio of the Reynolds number ${\rm R_e}$ and the P${\rm \acute e}$clet number ${\rm P_e}$ such that ${\rm S_c\!=\!P_e/R_e}$. The Reynolds number is evaluated to be ${\rm R_e}\!=\!Vd/\nu\!=\!5\!\times\!10^{-5}$ because the velocity, diameter and kinematic viscosity are assumed to be $V\!=\!50\!\times\! 10^{-6}~{\rm m/s}$, $d\!=\!500\!\times\! 10^{-6}~{\rm m}$ and $\nu\!=\!1\!\times\!10^{-4}~{\rm m^2/s}$, respectively. The P${\rm \acute e}$clet number is similarly evaluated to be ${\rm P_e}\!=\!Vd/D_{\rm dif}\!\simeq\!10$ for $D_{\rm dif}\!=\!5.3\!\times\! 10^{-10}~{\rm m^2/s}$. If we assume that $D_{\rm dif}\!=\!4\!\times\! 10^{-11}~{\rm m^2/s}(=\!2D_{e,0} (n_T{\it \Delta}t_0 \beta_0) ^2)$, we obtain ${\rm P_e}\!\simeq\!100$, which is closer to the estimate of $10^2\!-\!10^3$ given in Ref. \cite{Goldstein-etal-PRL2008}.
Therefore, we obtain ${\rm S_c}\!\simeq\!2\!\times\!10^5$ for $D_{\rm dif}\!=\!5.3\!\times\! 10^{-10}~{\rm m^2/s}$ and ${\rm S_c}\!\simeq\!2\!\times\!10^6$ for $D_{\rm dif}\!=\!4\!\times\! 10^{-11}~{\rm m^2/s}$. Note also that ${\rm S_c}$ can be obtained directly from its definition, ${\rm S_c}\!=\!\nu/D_{\rm dif}$. Thus, the relatively large ${\rm S_c}$ implies that the viscosity force is larger than the diffusion force, which is activated by thermal fluctuations.
Therefore, the effect of thermal fluctuations is relatively small compared not only to the advection term \cite{Goldstein-etal-PRL2008,Goldstaein-etal-PNAS2008} but also to the viscosity term. Nevertheless, the peaks in the velocity distribution essentially originate from this small thermal fluctuation effect.

\subsection{Lattice size dependence\label{lat-size-dep}}
\begin{figure}[h]
\begin{center}
\includegraphics[width=9.5cm]{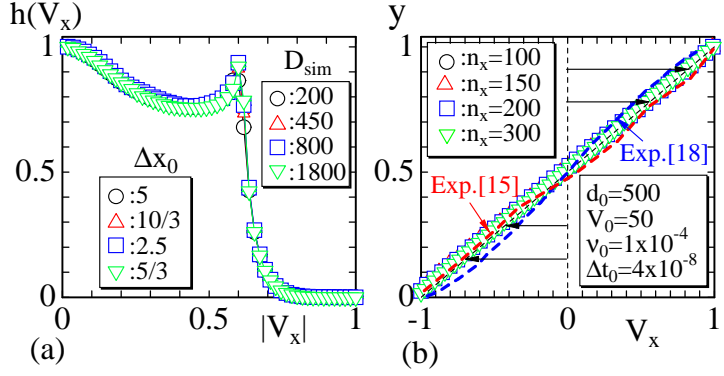}
\caption{(a) Distributions $h(V_x)$ obtained on lattices with sizes ranging from $n_X\!=\!100$ to $n_X\!=\!300$. (b) The dependence of $V_x$ on $y$. The parameters $D_{\rm sim}$ and ${\it \Delta} x_0$ are varied in accordance with Eq. (\ref{param-changes}), and $D_{\rm sim}$, ${\it \Delta} x_0$ and the other parameters $d_0$, $V_0$, $\nu_0$, and ${\it \Delta}t_0$ shown in (a) and (b) are all expressed in the simulation units.  Dashed lines in (b) denoted by Exp. are normalized experimental data reported in Refs. \cite{Kikuchi-Mochizuki-PlosOne2015,Goldstaein-etal-JFM2010}.}\label{fig-8}
\end{center}
\end{figure}
In this subsection, we show that the simulation results are independent of the lattice size $n_X$.
As mentioned in Section \ref{invariance}, we need to change not only $n_X$ but also $D_{\rm sim}$, to observe this dependence.
Figure \ref{fig-8}(a) shows the histograms $h(V_x)$ with respect to $|V_x|$ obtained on lattices with sizes ranging from $n_X\!=\!100$ to $n_X\!=\!300$. In these simulations, the parameters $D_{\rm sim}$ and ${\it \Delta} x_0$ are scaled to $\gamma^2D_{\rm sim}$ and $\gamma^{-1}{\it \Delta} x_0$, as indicated in Eq. (\ref{param-changes}). The results remain unchanged when the lattice size is modified from $n_X(=\!100)$ to $\gamma n_X$, where $1\!\leq\! \gamma\!\leq\! 3$ (see Eq. (\ref{scale-changes})). We start with $\gamma\!=\!1$ for $D_{\rm sim}\!=\!\gamma^2 200$ and ${\it \Delta} x_0\!=\! \gamma^{-1} 5~[{\rm \alpha m}]$ with ${\it \Delta} t_0\!=\!4\times 10^{-8}~{\rm \beta s}$. The parameters $D_{\rm sim}$ and ${\it \Delta} t_0$ can also be replaced with $D_{\rm sim}\!=\!\gamma^2 400$ and ${\it \Delta} t_0\!=\!8\times 10^{-8}~{\rm \beta s}$, which are identical to those plotted in Figs. \ref{fig-7}(a) and \ref{fig-7}(b). The reason $D_{\rm sim}\!=\!\gamma^2 400$ and ${\it \Delta} t_0\!=\!8\times 10^{-8}~{\rm \beta s}$ are replaced with $D_{\rm sim}\!=\!\gamma^2 200$ and ${\it \Delta} t_0\!=\!4\times 10^{-8}~{\rm \beta s}$ is that if we start with ${\it \Delta} t_0\!=\!8\times 10^{-8}~{\rm \beta s}$, the simulation does not converge for the case of $\gamma\!=\!3$.

The height of the histogram $h(V_x)$ is normalized such that the maximum height is equal to 1 for each $D_{\rm sim}$, as in Fig. \ref{fig-7}(a), while the horizontal axes $|V_x|$ for all $D_{\rm sim}$ are normalized using a constant value equal to the maximum $|V_x|$ for $D_{\rm sim}\!=\!200$ and a lattice size of $n_X\!=\!100$ satisfying $h(V_x)\!\not=\!0$.

We find that $h(V_x)$ with respect to $|V_x|$ is almost independent of either $n_X$ or ${\it \Delta} x$. This finding supports not only the correctness of statement (B) but also the ${\it \Delta} x$ independence of the results. In fact, if $h(V_x)$ with respect to $|V_x|$ did depend on ${\it \Delta} x$, we would need to consider either statement (B) to be incorrect or the results to depend on ${\it \Delta} x$.

Additionally, the dependence of $V_x$ on $y$, plotted in Fig. \ref{fig-8}(b), is almost linear, and this linear behavior is the same as that of the trivial solution in Eq. (\ref{trivial-sol}).
We should note that this linear behavior is different from the previously experimentally observed  behaviors \cite{Kikuchi-Mochizuki-PlosOne2015,Goldstaein-etal-JFM2010}.  The reason for this deviation is considered to be the simplification of the model, as stated in the Introduction and in Section \ref{discretization}.
The $y$ dependence of $V_x$ is found simply by averaging $V_x$ from the convergent configurations using Eq. (\ref{discrete-velocity}).
Thus, the nontrivial behavior of two distinct peaks in $h(V_x)$ is not always reflected in the dependence of $V_x$ on $y$. The parameters used in the simulations are shown in Fig. \ref{fig-8}(b).

\subsection{Discrete time-step dependence\label{time-step-dep}}
\begin{figure}[ht]
\begin{center}
\includegraphics[width=9.5cm]{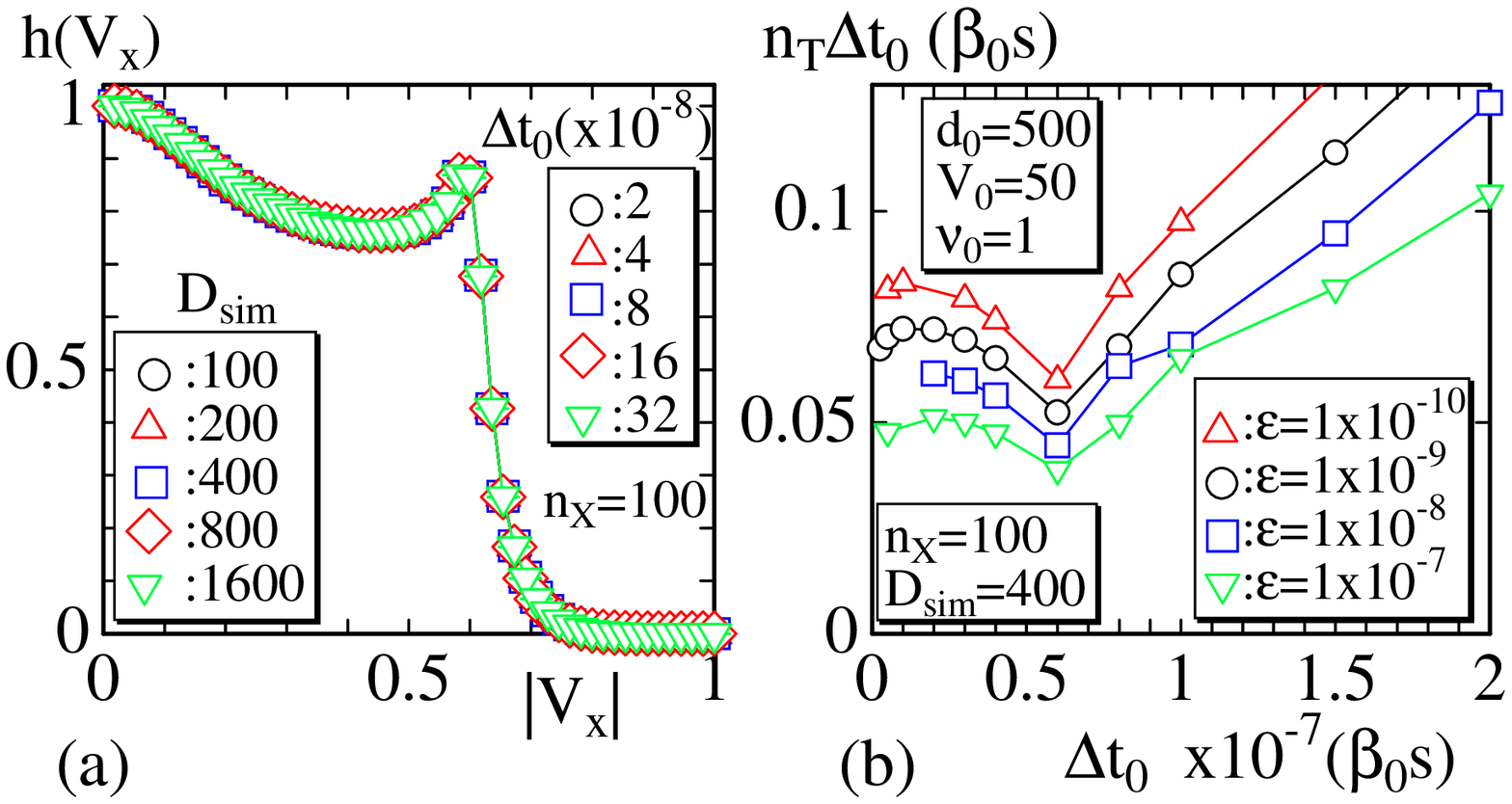}
\caption{(a) Distributions $h(V_x)$ with respect to $|V_x|$ for different combinations of ${\it \Delta} t_0$ and $D_{\rm sim}$. (b) $n_T{\it \Delta} t_0$ vs. ${\it \Delta} t_0$. In (a), the parameters ${\it \Delta} t_0$ and $D_{\rm sim}$ are scaled to $\delta^{-1}{\it \Delta} t_0$ and $\delta^{-1}D_{\rm sim}$, where ${\it \Delta} t_0$ is varied in the range $2\!\times\! 10^{-8}\!\leq\!{\it \Delta} t_0\!\leq\! 32\!\times\! 10^{-8}$ with the simulation units (${\rm 1~\beta_0 s}\!=\!{\rm 0.1~s}$).}
\label{fig-9}
\end{center}
\end{figure}
In this subsection, the dependence of the results on ${\it \Delta} t$ is checked, as mentioned in Section \ref{invariance}. This subsection contributes to the numerical proof of statement (B). There are two possible origins, $\tau$ and $n_T$, for the change in ${\it \Delta} t$, as described in Eq. (\ref{sim-parameters}). However, in sharp contrast to the case of the diameter $d$ for ${\it \Delta}x$, the relaxation time $\tau$ is not clear. Nevertheless, to check the scaling properties in Eq. (\ref{param-changes}), it is sufficient to observe the dependence of the results on ${\it \Delta} t$ or $\delta$. The normalization of the histograms $h(V_x)$ with respect to $|V_x|$ is defined in the same way as in Fig. \ref{fig-8}(a).

In Fig. \ref{fig-9}(a), the histograms $h(V_x)$ of velocity vs. $|V_x|$ are plotted. The parameters $D_{\rm sim}$ and ${\it \Delta} t_0$ are scaled to $\delta^{-1}D_{\rm sim}$ and $\delta^{-1}{\it \Delta} t_0$.
The results are independent of ${\it \Delta}t_0$, which is varied in the range $2\!\times\! 10^{-8}\leq{\it \Delta} t_0\!\leq\! 32\!\times\! 10^{-8}$ in the simulation units (${\rm 1~\beta_0 s}\!=\!{\rm 0.1~s})$. This range of ${\it \Delta} t_0$ corresponds to a $\delta$ range of $1\leq\delta\leq 16$. Thus, the results plotted in Fig. \ref{fig-9}(a) confirm that the corresponding scaling properties, such as $\delta^{-1}D_{\rm sim}$ and $\delta^{-1}{\it \Delta} t_0$, are correct.
This completes the numerical verification of statement (B).

Finally, in this subsection, we check whether $n_T{\it \Delta} t_0$ depends on the $\varepsilon$ used for the convergence criteria in Eqs. (\ref{conv-time-step}) and (\ref{conv-for-psi}), which has previously been fixed to $\varepsilon\!=\! 1\!\times\! 10^{-9}$ for all simulations.
Here, the value is additionally set to $\varepsilon\!=\! 1\!\times\! 10^{-7}$, $\varepsilon\!=\! 1\!\times\! 10^{-8}$ and $\varepsilon\!=\! 1\!\times\! 10^{-10}$ to assess the dependence of $n_T{\it \Delta} t_0$ on $\varepsilon$. The results shown in Fig. \ref{fig-9}(b) indicate that $n_T{\it \Delta} t_0$ is roughly independent of $\varepsilon$ for sufficiently small ${\it \Delta} t_0$.

\subsection{Dependence on physical parameters\label{additional}}
In this subsection, we demonstrate how to use statement (B) to discuss the dependence of the normalized velocity distribution on the physical parameters $\nu_e$, $V_e$ and $d_e$.
From the perspective of statement (B), the purpose of the simulations in Section \ref {velocity-distribution} is to find $S_0=(\nu_0,  V_0,   D_0, {\it \Delta} x_0, {\it \Delta} t_0)$ with a suitable $D_0$ and the set of parameters listed in Table \ref{table-3}. Indeed, from the simulation results, we find that $D_0\!=\!400$ in the second line of Table \ref{table-4} is suitable because the shape of $h(V_x)$ in Fig. \ref{fig-7}(a) is relatively close to the experimental data reported in Refs. \cite{Mustacich-Ware-PRL1974,Mustacich-Ware-BJ1976,Mustacich-Ware-BJ1977}.  
Since $|V_x|$ in Fig. \ref{fig-7}(a) is normalized, the position of the second peak can only be compared to those in Figs. \ref{fig-2}(a) and \ref{fig-2}(b). The peak position of the result of $D_{\rm sim}\!=\!400$ is approximately 0.6, while that of Fig. \ref{fig-2}(b) is approximately 0.5; the deviation from the peak in Fig. \ref{fig-2}(a) is clearly larger. However, the positions of the second peaks in Figs. \ref{fig-2}(a) and \ref{fig-2}(b) move to the right if the high-frequency part, where the intensity is close to zero, is removed. Although the relative intensity of the second peak in the numerical data is higher than those in Figs. \ref{fig-2}(a) \ref{fig-2}(b), we consider that the existence of the second peak is clearly reproduced by this simplified 2D model.
Thus, using the parameters $S_0$, we can perform simulations of ${\rm Exp}(E_e)$ characterized by $E_e$, which is different from $E_{e,0}$.

The parameters are shown in Table \ref{table-5}, where $\nu_e$, $V_e$, and $d_e$ are the elements of the experimental data $E_{(i)}$ $(i=1,2,3)$. $E_{(1)}$, $E_{(2)}$, and $E_{(3)}$ are different from $E_{e,0}$ in Table \ref{table-2} only in terms of $\nu_e$, $V_e$, and $d_e$, respectively (indicated by underlines). 
Note that ${\nu_e}/{\nu_{e,0}}\!=\!2$, ${V_e}/{V_{e,0}}\!=\!2$, and ${d_e}/{d_{e,0}}\!=\!2$ imply that ${\nu_e}\!=\!2{\nu_{e,0}}\!=\!2\times10^{-4}\; {\rm m^2/s}$, ${V_e}\!=\!2{V_{e,0}}\!=\!100\; {\rm \mu m/s}$, and ${d_e}\!=\!2{d_{e,0}}\!=\!1\; {\rm mm}$. All of these values are meaningful in the engineering viewpoint, because these parameters in plant cells such as \textit{Chara corallina} and \textit{Nitella flexilis} are not always uniquely determined but distributed around the values ${\nu_{e,0}}$, ${V_{e,0}}$, and ${d_{e,0}}$ in Table \ref{table-2} depending on their size \cite{Kamiya-Kuroda-1973}. 
\begin{table}[htb]
\caption{Experimental data $E_{(i)}$ $(i\!=\!1,2,3)$, ${\tau_{e}}$, ${D_e}$, and the corresponding parameters $\alpha$, $\beta$, $\gamma$, $\delta$ and $D_{\rm sim}$ assumed for the simulation parameters $S_{e(i)}$ $(i\!=\!1,2,3)$. These parameters are determined by the ratio $D_{\rm sim}/D_0$, where $D_0\!=\!400$ from the second line of Table \ref{table-4}.   }
\label{table-5}
\begin{center}
 \begin{tabular}{cccccccccccccccccccccc}
 \hline
$E_{(i)}$  & $\frac{\nu_e}{\nu_{e,0}}$ && $\frac{V_e}{V_{e,0}}$ && $\frac{d_e}{d_{e,0}}$&& $\frac{\tau_{e}}{\tau_{e,0}}$ && $\frac{D_e}{D_{e,0}}$ &&$\frac{\alpha}{\alpha_0}$ &&$\frac{\beta}{\beta_0}$ &&$\frac{\gamma}{\gamma_0}$ &&$\frac{\delta}{\delta_0}$ && $\frac{D_{\rm sim}}{D_0}$    \\
 \hline
 (1) & $\underline{2}$       &&  1   &&   1    &&  $\frac{1}{2}$  && 2    && 2    && 2     && $\frac{1}{2}$  && $\frac{1}{4}$  && 4   \\
 (2) &1       &&  $\underline{2}$   &&   1    &&  1    && 1  && $\frac{1}{2}$       && $\frac{1}{4}$ && $2$    && $4$    && $\frac{1}{16}$   \\
 (3) &1       &&  1   &&  $\underline{2}$     &&  4   && $\frac{1}{4}$   && 1       && 1     && $2$  && $4$    && $\frac{1}{4}$  \\
 \hline
\end{tabular}
\end{center}
\end{table}

We assume that the macroscopic relaxation time $\tau_e$ in Eq. (\ref{ESS-formula}) is proportional to the inverse kinematic viscosity $\nu_e^{-1}$ \cite{Zaochik-ETFS1997} and the area $A_e$ as follows:
\begin{eqnarray}
\label{viscosity}
\tau_e\sim A_e/\nu_e.
\end{eqnarray}
Using this relation, we obtain the ratio $\tau_e/\tau_{e,0}$ in Table \ref{table-5}.
Consequently, from Eq. (\ref{ESS-formula}), we have
\begin{eqnarray}
\label{ESS-formula-2}
D_e\sim\frac{k_BT}{\mu a}\tau_{e}^{-2}\sim d_e^{-2} \nu_e, 
\end{eqnarray}
where $A_e$ is replaced with $d_e^2$, the viscosity $\mu$ is proportional to $\nu_e$, and the temperature is assumed to be constant.

We will now explain how to obtain the values of ${D_{\rm sim}}/{D_0}$ via statement (B).
Since ${\rm Exp}(E_{e,0})$ is simulated with $S_0\!=\!(\nu_0,  V_0,  D_0, {\it \Delta} x_0, {\it \Delta} t_0)$, from statement (B), we have
\begin{eqnarray}
\label{theorem-params-1}
\begin{split}
&(\nu_0,  V_0,  D_0, {\it \Delta} x_0, {\it \Delta} t_0) \\
&=(\alpha_0^{-2}\beta_0\nu_{e,0},  \alpha_0^{-1} \beta_0 V_{e,0},    \alpha_0^{-2}\beta_0^{3}\gamma_0^2\delta_0^{-1} D_{e,0}, \\
&\quad \alpha_0^{-1}\gamma_0^{-1}{\it \Delta} x_{e,0}, \beta_0^{-1}\delta_0^{-1}{\it \Delta} t_{e,0}),
\end{split}
\end{eqnarray}
using the parameters $\alpha_0$, $\beta_0$, $\gamma_0$, and $\delta_0$ for the scale transformation between $S_0$ and $S_{e,0}$. This is the assumption component, or the trivial case of statement (B), and is exactly the same as Eq. (\ref{theorem-params-0}).
We also have
\begin{eqnarray}
\label{theorem-params-2}
\begin{split}
&(\nu_0,  V_0,  D_{\rm sim}, {\it \Delta} x_0, {\it \Delta} t_0) \\
&=(\alpha^{-2}\beta\nu_e,  \alpha^{-1} \beta V_e,    \alpha^{-2}\beta^{3}\gamma^2\delta^{-1} D_{e} \\
&\quad \alpha^{-1}\gamma^{-1}{\it \Delta} x_e, \beta^{-1}\delta^{-1}{\it \Delta} t_e),
\end{split}
\end{eqnarray}
using the parameters $\alpha$, $\beta$, $\gamma$, and $\delta$ for the scale transformation between $S_0$ and $S_{e}$. Therefore, from Table \ref{table-5}, we have
\begin{eqnarray}
\label{eg-1}
\begin{split}
&\alpha_0^{-2}\beta_0=2\alpha^{-2}\beta, \quad  \alpha_0^{-1} \beta_0=\alpha^{-1} \beta      \quad {\rm for} \; E_{(1)},  \\
&\alpha_0^{-2}\beta_0=\alpha^{-2}\beta, \quad  \alpha_0^{-1} \beta_0=2\alpha^{-1} \beta      \quad {\rm for} \; E_{(2)},  \\
&\alpha_0^{-2}\beta_0=\alpha^{-2}\beta, \quad  \alpha_0^{-1} \beta_0=\alpha^{-1} \beta      \quad {\rm for} \; E_{(3)}, 
 \end{split}
\end{eqnarray}
which imply
\begin{eqnarray}
\label{eg-2}
\begin{split}
&\alpha/\alpha_0=2,\quad \beta/\beta_0=2      \quad {\rm for} \; E_{(1)},  \\
&\alpha/\alpha_0=1/2,\quad \beta/\beta_0=1/4     \quad {\rm for} \; E_{(2)},  \\
&\alpha/\alpha_0=1, \quad \beta/\beta_0=1      \quad {\rm for} \; E_{(3)}.  \\
 \end{split}
\end{eqnarray}
From Eqs. (\ref{gamma}) and (\ref{delta}), we have   $\gamma/\gamma_0$ and $\delta/\delta_0$ as listed in Table \ref{table-5} for $E_{(1)}$ and $E_{(2)}$.
Thus, we obtain ${D_{\rm sim}}/D_0$ using these values of $\alpha/\alpha_0$, $\beta/\beta_0$, $\gamma/\gamma_0$, and $\delta/\delta_0$ and by Eqs. (\ref{theorem-params-1}) and (\ref{theorem-params-2}), as follows:
\begin{eqnarray}
\label{D_sim}
\frac{D_{\rm sim}}{D_0}= \left(\frac{\alpha}{\alpha_0}\right)^{-2} \left(\frac{\beta}{\beta_0}\right)^{3}\left(\frac{\gamma}{\gamma_0}\right)^{2}\left(\frac{\delta}{\delta_0}\right)^{-1}\frac{D_e}{D_{e,0}}.
\end{eqnarray}
Therefore, the experimental data corresponding to ${\rm Exp}(E_{(i)})$ $(i=1,2,3)$ can be simulated with \begin{eqnarray}
\label{D_sim_value}
D_{\rm sim}=1600\; (E_{(1)}),\quad 25\; (E_{(2)}),\quad 100\; (E_{(3)}) 
\end{eqnarray}
and with the parameters in Table \ref{table-3}. Thus, we expect the peak position for $E_{(1)}$ ($E_{(2)}$ or $E_{(3)}$) to move to the left (right) of the peak position for $E_{e,0}$. Moreover, the simulation results for ${\rm Exp}(E_{(i)})$ $(i=1,2,3)$ are located between, or are only slightly different from, the curves in Figs. \ref{fig-7}(a) and \ref{fig-7}(b). This is why we regard the results in Section \ref{velocity-distribution} as our main results, which we emphasize in this subsection.
Note that the results are true only if the assumption regarding $\tau_e$ in Eq. (\ref{viscosity}) is true. It should also be noted that the results in Eq. (\ref{D_sim_value}) are qualitatively reasonable. Indeed, the Reynolds number ${\rm R_e}(=\!Vd/\nu)$ is decreased for $E_{(1)}$ and increased for $E_{(2)}$ and $E_{(3)}$, and consequently, the second peak position is expected to move in opposite directions depending on the variation of ${\rm R_e}$.

Finally, we must note the implications of statement (B) and its supporting analyses in this subsection. Our main result that {all qualitatively different simulation results can be obtained merely by varying the strength $D$ of the random Brownian force}, as stated in the Introduction, is limited in the sense that this is true only if the simulation results obtained with the parameter set $S_0$ correspond to ${\rm Exp}(E_{e,0})$. In fact, this assumption is not always exactly satisfied because the experimental data in Figs. \ref{fig-2}(a) and \ref{fig-2}(b) are not exactly the same as the simulation data for $D_{\rm sim}\!=\!400$ in Fig. \ref{fig-7}(a). 
However, for the second peak position, as mentioned above, these experimental and simulation data are almost the same; therefore, statement (B) indicates that the second peaks of all the experimental normalized velocity distributions corresponding to different $(\nu_e,V_e,D_e)$, such as those in Table \ref{table-4}, are identical to or located between the peaks in Fig. \ref{fig-7}(a).

\section{Summary and conclusion\label{summary}}
In this paper, we study the flow fields associated with protoplasmic streaming in plant cells such as \textit{Chara corallina} and \textit{Nitella flexilis} by means of the stochastic or Langevin Navier-Stokes (NS) equation, which is a 2D equation for describing incompressible viscous flows with random Brownian forces. The study focuses on the experimentally observed distribution of the velocity along the flow direction, which exhibits two distinct peaks. 
To clearly illustrate the role of the random Brownian force assumed in the NS equation, the computational model is simplified such that the twist of the flows is neglected and 2D Couette flow is assumed.

From a dimensional analysis of the Langevin NS equation, we find that the normalized velocity distribution depends only on the strength $D$ of the random Brownian force. This finding is numerically verified in detail in Section \ref{velocity-distribution}. Furthermore, in Section \ref{additional}, we extract the reasonable finding that the position of the second peak moves to the right or left in accordance with the variation in the physical parameters, i.e., the kinematic viscosity, diameter and boundary velocity. If the kinematic viscosity is decreased, for example, then the peak position is expected to move to the right or a higher-velocity region. This phenomenon is consistent with the intuitive understanding of the Reynolds number in the sense that a decrease in the kinematic viscosity is equivalent to an increase in velocity.

The results in this paper imply that the spiral flow and the 3D nature of real protoplasmic streaming are not essential for the emergence of the two peaks in the velocity distribution, although the shapes of the peaks are expected to be influenced by these experimentally observed characteristics. Rather, the random Brownian forces, represented by Gaussian random numbers, are confirmed to be the origin of the peaks. 

As noted in the final part of Section \ref{velocity-distribution}, the dependence of $V_x$ on $y$ is almost linear and is slightly different from the experimentally reported results. One reason for this deviation is that compared with real flows, the simulation model used in this paper is simplified in many respects. In particular, this simulation model is a 2D model, and the twist of the flows and the interactions between the fluid and biological materials are neglected, as mentioned above. These neglected components should be included in the model  for further fluid mechanics studies on protoplasmic streaming.

\section{Data Availability Statement}
The data that support the findings of this study are available from the corresponding author upon reasonable request.

\acknowledgements
Author H.K. acknowledges Dr. Kazuhiko Mitsuhashi for reminding him of this interesting field. This work is supported in part by a Collaborative Research Project of the Institute of Fluid Science (IFS), Tohoku University, and it is also supported in part by a Collaborative Research Project of the National Institute of Technology (KOSEN), Sendai College. H.K. acknowledges Prof. Jean-Yves Cavaille of INSA Lyon for the support of the IFS project and encouragement. Author V.E. acknowledges president Dr. Hiroshi Fukumura of Sendai KOSEN for the warm hospitality provided during a four-month stay from 2019 to 2020, which was supported in part by JSPS KAKENHI Grant No. JP17K05149.

\appendix

\section{Proof of Statement (B) in Section \ref{uniqueness}} \label{append-A}
Statement (B) in Section \ref{uniqueness} is as follows: 
\noindent
\begin{enumerate}
\item[(B)]
Let $E_{e,0}=(\nu_{e,0},  V_{e,0},  d_{e,0})$ be a set of parameters that characterize ${\rm Exp}(E_{e,0})$, and let $S_0$ be a set of parameters given by $S_0=(\nu_0,  V_0,   D_0, {\it \Delta} x_0, {\it \Delta} t_0)$. In this situation, if ${\rm Exp}(E_{e,0})\!\simeq\! S_0$, then for any experimental data ${\rm Exp}(E_e)$, with $E_e\!=\!(\nu_e,  V_e,  d_e)$ and $D_e$, and for any given set of $(n_X, n_T)$, there exists a unique $D_{\rm sim}$ such that ${\rm Exp}(E_e)\simeq (\nu_0,   V_0,    D_{\rm sim}, {{\it \Delta} x_0}, {{\it \Delta} t_0})$.
\end{enumerate}

To prove statement (B), we first fix the parameters $\alpha$ and $\beta$ using the parameter sets $S_0$ and $E_e$ such that
\begin{eqnarray}
\label{alpha-beta-1}
\alpha=\frac{\nu_e}{\nu_0}\frac{V_0}{V_e}, \quad \beta=\frac{\nu_e}{\nu_0}\left(\frac{V_0}{V_e}\right)^{\!\!2}.  
\end{eqnarray}
Indeed, from the expressions $\nu_e~[{\rm m^2}/{s}]\!=\!\nu_e \alpha^{-2}\beta~[{\rm (\alpha m)^2}/{\rm\beta s}]$ and $V_e~[{\rm m}/{\rm s}]\!=\!V_e \alpha^{-1}\beta~[{\rm \alpha m}/{\rm \beta s}]$, we have $\nu_0\!=\!\nu_e \alpha^{-2}\beta$ and $V_0\!=\!V_e \alpha^{-1}\beta$, which lead to Eq. (\ref{alpha-beta-1}). Note that $\alpha$ and $\beta$ in Eq. (\ref{alpha-beta-1}) correspond to the unit transformation between $E_e$ and $S_0$ and are not always identical to $\alpha_0$ and $\beta_0$ for the unit transformation between $E_{e,0}$ and $S_0$.

The assumption ${\rm Exp}(E_{e,0})\!\simeq\! S_0$ implies that there exist parameters $\alpha_0$, $\beta_0$, $\gamma_0$ and $\delta_0$ for the scale transformations ${\rm m \!\to\! \alpha_0 m}$, ${\rm s \!\to\! \beta_0 s}$, ${ n_{X} \!\to\! \gamma_0 n_{X}}$, and ${ n_{T} \!\to\! \delta_0 n_{T}}$ such that
\begin{eqnarray}
\label{theorem-params-0}
\begin{split}
&(\nu_0,  V_0,  D_0, {\it \Delta} x_0, {\it \Delta} t_0) \\
&=(\alpha_0^{-2}\beta_0\nu_{e,0},  \alpha_0^{-1} \beta_0 V_{e,0},    \alpha_0^{-2}\beta_0^{3}\gamma_0^2\delta_0^{-1} D_{e,0}, \\
&\quad \alpha_0^{-1}\gamma_0^{-1}{\it \Delta} x_{e,0}, \beta_0^{-1}\delta_0^{-1}{\it \Delta} t_{e,0}).
\end{split}
\end{eqnarray}

The parameter $D_e$ is assumed to be given in addition to $E_e$ for ${\rm Exp}(E_e)$, and it is not always identical to $D_{e,0}$.
The parameters ${\it \Delta}x_e$ and $ {\it \Delta}t_e$ are defined as
\begin{eqnarray}
\label{alpha-beta-2}
 {\it \Delta}x_e=\gamma {{\it \Delta}x_0}\alpha, \quad {\it \Delta}t_e= \delta {\it \Delta}t_0\beta,
\end{eqnarray}
where ${\it \Delta}x_0$ and ${\it \Delta}t_0$ are given by
\begin{eqnarray}
\label{lattice-spacing}
{\it \Delta}x_0=\frac{d_{e}}{\gamma n_{X}}\alpha^{-1}=\frac{d_{e}}{\gamma n_{X}}\frac{\nu_0}{\nu_{e}}\frac{V_{e}}{V_0} \; [{\rm \alpha m}],
\end{eqnarray}
\begin{eqnarray}
\label{time-spacing}
{\it \Delta}t_0=\frac{\tau_{e}}{\delta n_{T}}\beta^{-1}=\frac{\tau_{e}}{\delta n_{T}}\frac{\nu_0}{\nu_{e}}\left(\frac{V_{e}}{V_0}\right)^2 \; [{\rm \beta s}].
\end{eqnarray}
The parameter $\gamma/\gamma_0$ is obtained from the constraint that ${\it \Delta}x_0$ in Eq. (\ref{lattice-spacing}), expressed in units of ${\rm \alpha m}$ for the simulation of ${\rm Exp}(E_e)$ and expressed in units of ${\rm \alpha_0 m}$ for the simulation of ${\rm Exp}(E_{e,0})$, is identical to ${\it \Delta}x_0$; thus, we have
\begin{eqnarray}
\label{gamma}
\frac{\gamma}{\gamma_0}=\frac{d_e}{d_{e,0}}  \frac{\nu_{e,0}}{\nu_{e}}\frac{V_e}{V_{e,0}}.
\end{eqnarray}
This expression indicates that $\gamma$ is uniquely determined because the parameters that appear in this expression are already uniquely given.
The uniqueness of the parameter $\delta$ can be understood from a similar expression obtained through the same procedure,
\begin{eqnarray}
\label{delta}
\frac{\delta}{\delta_0}=\frac{\tau_e}{\tau_{e,0}}  \frac{\nu_{e,0}}{\nu_{e}}\left(\frac{V_e}{V_{e,0}}\right)^2,
\end{eqnarray}
although we must assume that the macroscopic relaxation time $\tau_e$ is a well-defined quantity in the target experiments corresponding to $E_{e,0}$ and $E_{e}$.

Using $D_e$, ${\it \Delta}x_e$ and ${\it \Delta}t_e$, we define a set of parameters $S_e$ such that
\begin{eqnarray}
\label{S-e-param}
S_e=(\nu_e, V_e,    D_e, {\it \Delta} x_e, {\it \Delta} t_e).
\end{eqnarray}

The strength $D_{\rm sim}$ of the random force is fixed to
\begin{eqnarray}
\label{D-sim}
 D_{\rm sim}=\alpha^{-2}\beta^{3} \gamma^2\delta^{-1}D_e,
\end{eqnarray}
which is unique because the quantities on the RHS are all uniquely given.
Thus, we have proven that
\begin{eqnarray}
\label{theorem-params}
\begin{split}
&(\nu_0,  V_0,  D_{\rm sim}, {\it \Delta} x_0, {\it \Delta} t_0) \\
&=(\alpha^{-2}\beta\nu_e,  \alpha^{-1} \beta V_e,    \alpha^{-2}\beta^{3}\gamma^2\delta^{-1} D_e, \\
&\quad \alpha^{-1}\gamma^{-1}{\it \Delta} x_e, \beta^{-1}\delta^{-1}{\it \Delta} t_e).
\end{split}
\end{eqnarray}
The relation in Eq. (\ref{theorem-params}) implies that $S_e\equiv (\nu_0,  V_0,  D_{\rm sim}, {\it \Delta} x_0, {\it \Delta} t_0)$, which means that ${\rm Exp}(E_e)$ can be simulated by $(\nu_0,  V_0,  D_{\rm sim}, {\it \Delta} x_0, {\it \Delta} t_0)$ from statement (A), thus concluding the proof of statement (B).

In terms of its rigor, this proof is insufficient because the macroscopic relaxation time $\tau_e$ is not always explicitly given in actual experimental data corresponding to $E_e$. In such a case, the expressions in Eqs. (\ref{time-spacing}) and (\ref{delta}) are meaningless. Therefore, this component is studied numerically in the results section. The problem to be numerically clarified is whether the scaling properties expressed in Eq. (\ref{theorem-params}) are correct for the cases of $\alpha\!=\!\beta\!=\!1$, $\gamma\!\not=\!1$, and $\delta\!\not=\!1$. It is sufficient to check only the case of $\delta\!\not=\!1$; however, the case of $\gamma\!\not=\!1$ will also be checked. It is numerically shown in Sec. \ref{results} shown that these scaling properties are correct. Thus, we assume that statement (B) is correct.

The lattice spacing ${\it \Delta}x_0$ in $S_0$ used in Eq. (\ref{alpha-beta-2}) is defined in terms of the experimental diameter $d_{e,0} $~[m] and the lattice size $n_{X0}(=\!\gamma_0 n_X)$, similar to Eq. (\ref{lattice-spacing}), such that
\begin{eqnarray}
\label{lattice-spacing-2}
{\it \Delta}x_0=\frac{d_{e,0}}{n_{X0}}\alpha_0^{-1}=\frac{d_{e,0}}{n_{X,0}}\frac{\nu_0}{\nu_{e,0}}\frac{V_{e,0}}{V_0} \; [{\rm \alpha_0 m}]
\end{eqnarray}
in the simulation units for ${\rm Exp}(E_{e,0})$. Since the diameter $d_{e,0}$ appears in ${\it \Delta}x_0$, it is not explicitly included in $S_0$ and $S_e$.
The parameter ${\it \Delta} x_{e}$ on the RHS of Eq. (\ref{theorem-params}) is not experimental and is simply defined by Eq. (\ref{alpha-beta-2}).
The final parameter in $S_e$, i.e., ${\it \Delta} t_e$, is also defined by Eq. (\ref{alpha-beta-2}).

The implications of statement (B) should be emphasized. The meaning of the equivalence between $S_e$ and $ (\nu_0,  V_0,  D_{\rm sim}, {\it \Delta} x_0, {\it \Delta} t_0)$, as expressed by $S_e\equiv (\nu_0,  V_0,  D_{\rm sim}, {\it \Delta} x_0, {\it \Delta} t_0)$, is that any experimental data ${\rm Exp}(E_e)$ characterized by the parameter $E_e$ can be simulated with a single set of parameters $S_0=(\nu_0,  V_0,   D_{0}, {\it \Delta} x_0, {\it \Delta} t_0)$ if $D_0$ is replaced with $D_{\rm sim}$. To simulate other experimental data ${\rm Exp}(E_{e}^\prime)$, it is sufficient to replace $D_{\rm sim}$ with $D_{\rm sim}^\prime$ such that $S_e^\prime\equiv (\nu_0,  V_0,  D_{\rm sim}^\prime, {\it \Delta} x_0, {\it \Delta} t_0)$.

$S_e\equiv (\nu_0,   V_0,    D_{\rm sim}, {{\it \Delta} x_0}, {{\it \Delta} t_0})$ simply implies that the simulation results with parameter set $S_e$ are identical to those with parameter set $(\nu_0,   V_0,    D_{\rm sim}, {{\it \Delta} x_0}, {{\it \Delta} t_0})$; hence, this equivalence does not always imply that the real experimental data characterized by $E_e$ are exactly the same as the simulation results obtained with $(\nu_0,   V_0,    D_{\rm sim}, {{\it \Delta} x_0}, {{\it \Delta} t_0})$. The latter problem is related to the fundamental problem of whether the Langevin NS simulation can successfully simulate real physical flows. In this paper, we assume that it can; this is the implication of the assumption that ${\rm Exp}(E_{e,0})\!\simeq\! S_0$. However, we emphasize that this assumption is true only because the experimentally observed peaks in the velocity distribution can be reproduced, which is the main result in this paper, as has been shown. 
Another implication of the assumption ${\rm Exp}(E_{e,0})\!\simeq\! S_0$ is that the set of parameters in $S_0$ is already given. Using the parameters in $S_0$ and $E_e$, we obtain $\alpha$ and $\beta$ via Eq. (\ref{alpha-beta-1}) for ${\rm Exp}(E_{e})$.

Although the kinematic viscosity coefficient $\nu$ appears in the NS equation given in Eq. (\ref{discrete-NS-eq}), statement (B) indicates that the simulation results depend only on $D$, which is understood from the original NS equation in Eq. (\ref{NS-eq-org}) for the velocity field without the pressure term:
\begin{eqnarray}
\label{NS-eq-org-2}
\frac{\partial {\vec V}}{\partial t}=-\left ({\vec V}\cdot \nabla\right){\vec V}+\nu {\it \Delta} {\vec V} + \vec\eta(t).
\end{eqnarray}
This equation contains two parameters, $\nu$  and $D$,  in the second and final terms, respectively. The first term can be neglected for protoplasmic streaming; this term is irrelevant to the following discussion, although it is included in Eq. (\ref{NS-eq-org-2}).
If the final term $\vec \eta$ is not present, it is easy to confirm that the solution is
\begin{eqnarray}
\label{trivial-sol}
 {\vec V}= \left(\frac{2V_B}{d}y,0\right),
\end{eqnarray}
which satisfies the boundary conditions in Eq. (\ref{boundary-velocity}) and is independent of $\nu$.
Thus, the question is whether this solution is also expected to satisfy Eq. (\ref{NS-eq-org-2}) and to be independent of $\nu$ in the presence of $\vec\eta(t)$. Statement (B) indirectly answers this question and shows that the results depend only on $D$ in the presence of $\vec \eta$, although this statement does not always imply that the results are independent of $\nu$.

\end{document}